\documentclass[floatfix,superscriptaddress,author-year,twocolumn, showpacs, aip, pra,10pt]{revtex4-1}
\usepackage{amsmath}
\usepackage{natbib}
\usepackage{hyperref}
\usepackage{graphicx}
\usepackage{bbm}
\usepackage{amssymb}
\usepackage[english]{babel}
\usepackage{lmodern}
\usepackage[T1]{fontenc}
\usepackage{isomath}
\usepackage[normalem]{ulem}
\usepackage{dcolumn}
\usepackage{longtable}
\usepackage{color}

\graphicspath{{Figs/}}

\usepackage{amssymb}

\renewcommand{\vec}{\vectorsym}

\begin{document}

\title{Survey of hyperfine structure measurements in alkali atoms}

\author{Maria Allegrini}
\affiliation{Dipartimento di Fisica ``E. Fermi'', Universit\`a di Pisa, Largo B. Pontecorvo 3, 56127 Pisa, Italy}
\affiliation{Istituto Nanoscienze-CNR, Piazza S. Silvestro 12, I-56127 Pisa, Italy}
\author{Ennio Arimondo}
\affiliation{Dipartimento di Fisica ``E. Fermi'', Universit\`a di Pisa, Largo B. Pontecorvo 3, 56127 Pisa, Italy}
\affiliation{Istituto Nazionale di Ottica-CNR, Via G. Moruzzi 1, 56124 Pisa, Italy}
\author{Luis A. Orozco}
\affiliation{Joint Quantum Institute, Dept. Physics, Univ. Maryland and National Institute of Standards and Technology, College Park, MD 20742, USA}

\begin{abstract}
The spectroscopic hyperfine constants for all the alkali atoms are reported. For atoms from lithium to cesium, only the long lived atomic isotopes are examined. For francium, the measured data for nuclear ground states of all available isotopes are listed. All results obtained since the beginning of laser investigations  are presented, while for previous works the data of Arimondo  {\it et. al.} Rev. Mod. Phys. 49, 31 (1977) are recalled. Global analyses based on the scaling laws and on the hyperfine anomalies are performed.
\end{abstract}

\date{\today}

\maketitle
\section{Note}
In press J. Phys. Chem. Ref. Data.

\section{Introduction}
\label{sec:intro}
Atomic spectroscopy was a key element in the original development of quantum mechanics theory at the beginning of the twentieth century. Its exploratory role continued after the Second World War, when microwave sources, very stable radiofrequency generators, optical pumping, and later on lasers, entered into the laboratories. The atomic contributions to quantum electrodynamics, parity violation, and the present searches for variation of the fundamental constants and for dark matter tests are  noteworthy. In this scenery, the alkali atoms and their hyperfine splittings represent an important reference because of their rather simple energy level structure and  their relatively easy laboratory exploration. Furthermore, they offer an opportunity to verify new experimental tools for a direct comparison within a wide research community. Very precise hyperfine constants of the alkalis are required for a large variety of atomic physics and quantum simulation experiments. More accurate hyperfine structure measurements have also revitalized their use in studies and tests of nuclear physics and fundamental symmetries in nature. Formidable progress achieved by atomic physics calculations supporting and also stimulating   research on the above advanced topics has refined its tools on the alkali hyperfine data. \\  
\indent A complete collection of hyperfine constants for alkali atoms was published by~\cite{Arimondo1977} at the time when laser sources introduced high-resolution atomic spectroscopy. Since that time, new spectroscopy tools have been developed, and technological advances have produced extremely precise atomic measurements. This progress is the motivation of the present work. The most amazing example of  the combination of scientific and technological progress is the atomic fountain proposed by Zacharias  in 1953, unpublished but described by~\cite{Ramsey1956}. Although it does not operate for room temperature atoms, it is very successful for launching ultracold ones, as exploited for hyperfine measurements based on atomic clocks, (e.g., see  ~\cite{GuenaBize2014,Ovchinnikov2015}). Using such tool, the hyperfine ground states of rubidium and cesium are presently measured with such a precision that   small variations of the fundamental constants can be tested. Recently, the use of frequency combs to perform  absolute optical frequency measurements has provided alkali hyperfine values with a precision increased by a factor of up one thousand. In addition, some well assessed spectroscopy tools have been refined. For instance, in~\cite{Bayram2014}, the  detection of delayed quantum beats at the hyperfine transition frequencies is used to determine very precise hyperfine coupling constants in several cesium excited  states, for which the precision of other techniques suffers from short lifetimes.  These approaches have increased the precision for a large set of hyperfine  measurements.\\
\indent In another class of experiments, the hyperfine constants have been determined for excited states accessible only by  laser sources covering new spectral regions or by multiple laser excitations. The most spectacular example is the alkali Rydberg states investigated for hyperfine structure up to levels with principal quantum number $n\approx  70$. For completeness,  here we report a  third class  of hyperfine measurements of pre-laser times for alkali states not recently investigated. A very interesting example of this class is the ground state hyperfine structures of lithium and sodium, for which the precise atomic-beam investigations of 1973-1974 remain the reference point. Certainly, atomic fountain experiments applied to those atoms could yield a precision comparable to that of the atomic clocks.\\
\indent  This work presents a complete overview of the measured hyperfine constants for the alkali atoms in ground or singly excited electronic states. It enlarges or supersedes the recent reviews of~\cite{DasNatarajan2008,KiranKumar2014R,WilliamsHawkins2018} that report a limited data set for lighter alkalis. We also add data to the francium review of~\cite{Sansonetti2007}. The main target is to provide to interested experimentalists and theoreticians  the full set of hyperfine data in an easily accessible form. We have collected the hyperfine data of the stable alkali isotopes  published after the review of~\cite{Arimondo1977}. In order to give a  complete overview, our Tables include measurements already reported in that review  in all cases where new and more precise values are not available. We do not examine the unstable isotopes for the alkalis from lithium up to cesium as for them very limited data were recently published.  Instead we review the full spectrum of the francium nuclear ground state isotopes, because of the recent interest associated with this atom as a test of the nuclear structure owing to the large number of explored isotopes for this alkali. This review intends to cover the experimental investigations, while discussing  theoretical results only briefly. However, it should be mentioned here that  theoretical comparison has greatly progressed, and global analyses for a given alkali atom have produced a large set of theoretical values for the hyperfine constants.\\
\indent For the experimentalists, this overview might stimulate investigations of specific atomic states for which the precision remains low. As advanced spectroscopic techniques are usually tested only on a few states,  a large set of high precision data might boost the theoretical effort for  global analyses. 
Based on ~\cite{NISTDatabase} ``NIST atomic energy levels and spectra bibliographic database'' (NIST stands for National Institute of Standards and Technology, USA), we have examined all articles that to the best of our knowledge have been published so far. However, we have disregarded publications that do not target the hyperfine splittings. For publications by the same research group reporting subsequent measurements with increasing precision our Tables include only the most recent value.\\
\indent Section II  presents the experimental tools exploited to measure the hyperfine constants. At first the atomic sample is examined from vapor cells up to ultracold atomic clouds where the spatial and velocity confinement greatly increases the spectral resolution. In the following the experiments are classified within some broad categories allowing a connection to the precision reached in ground and excited states.  The core part of this review is Sec. III. It is composed of Tables reporting the hyperfine constants for each alkali isotope classified on the basis of the atomic state and for a given state in chronological order.   Before presenting the Tables, the basic theoretical concepts of the hyperfine interaction are briefly  recalled.  For each atom we discuss the main results and we mention large discrepancies, if any, between the data of a given state.  For Rb and Cs atoms, where more data are available, Section IV presents scaling laws vs the quantum number of the excited states. Such scaling is applied to determine or confirm the sign of the hyperfine constants for several states. The scaling law is applied also to the  $S$ states of francium isotopes for which few data are available. Hyperfine anomalies producing information on the nuclear structure are discussed, at least for the  states measured with reasonable  precision. A short Section concludes this review.
\section{Spectroscopic tools}
\subsection{Samples}
{\it Atomic beam} (AB). In an atomic beam atoms propagate along a given direction with a small spread in the orthogonal plane, see~\cite{Ramsey1956}. Usually, an exciting laser propagates perpendicularly to the beam propagation leading to a very small Doppler broadening. \\  
\indent{\it Vapor cell} (VC). In a glass/quartz cell the atoms are  in the vapour phase and their vapour pressure and the atomic density are controlled by the cell temperature.\\
\indent {\it Magneto-optical trap} (MOT). The combined action of laser cooling and magnetic field confinement produces dense atomic samples having a greatly reduced Doppler linewidth. Such samples allow to detect weak absorption features, as for highly excited states, and to perform experiments with long interaction times leading to increased precision. \\
\indent {\it Fountain} (FOUNT). In a fountain, atoms from a MOT are launched vertically by radiation pressure or a moving optical lattice, see~\cite{MetcalfBook}. Excitation and detection take place at the same vertical position, the first one at the launching time and the second after the parabolic motion. Very high parabolic evolutions are used in order to increase the interrogation time. This approach is applied  to the atomic clocks. \\
\indent {\it Optical Dipole Trap} (ODT) In the experiment by~~\cite{NeuznerRitter2015} a single $^{87}$Rb atom is trapped by an optical dipole trap created by a 2-D optical lattice. Cavity-enhanced state detection of the optical absorption produces a good signal-to-noise ratio even for a single atom. Light-shift correction is carefully applied.\\
 \indent {\it Thermoionic diode} (TD). In presence of  a weak electrical discharge in  a VC, the light excited  atoms are ionized by electron collisions. These ions diffuse into the space charge region of the diode, compensate partially the space charge, and increase the thermionic diode current, as described in \cite{HerrmannWeis1985}

\subsection{Spectroscopic Techniques}
The experimental techniques used to measure the reported hyperfine measurements are classified in the following. Several research groups have introduced a specific name for their technique. While our classification scheme is concise, detailed presentations of the techniques can be found in textbooks, such as~\cite{Ramsey1956,Kopfermann1958,FootBook,BudkerDemilleBook,InguscioFallaniBook}.\\
\indent {\it Coherent-control spectroscopy} (CCS). This saturated absorption spectroscopy is based on copropagating pump and probe acting on a three-level V system. Similar to saturated absorption spectroscopy in a two-level system and to the electromagnetic-induced transparency in a $\Lambda$  level scheme, a pump laser originates an atomic coherence in a branch of the V scheme. The absorption profile of a probe laser is modified. The frequency difference between pump and probe lasers furnishes  the  excited state energy splitting, as in~\cite{DasNatarajan2005} \\
\indent {\it Delayed detection} (DD). The natural linewidth of a spectroscopic resonance is reduced by monitoring the atom evolution for times longer than the spontaneous emission lifetime. This refinement was combined with other techniques, such as as laser-induced fluorescence by~\cite{ShimizuTakuma1987} or hyperfine quantum-beats by~\cite{DeechSeries1977,KristAndrae1977,YeiSieradzan1993}.\\
\indent{\it  Double-resonance optical
pumping} (DROP). A double-resonance optical excitation on a ladder  three-level scheme is applied. An increased signal-to-noise ratio is obtained by detecting the population of the ground state rather than the excited state one. In the copropagating geometry, one laser excites zero velocity atoms on the lower transition and a second laser is scanned in frequency, as in~\cite{MoonLeeSuh2009}.  In the counterpropaganting geometry, it is often  combined with electomagnetic-induced transparency.\\
\indent {\it Electromagnetic induced transparency} (EIT). This coupling/probe spectroscopy is based on the very narrow coherent feature produced in the absorption spectrum of  three-level $\Lambda$ or ladder systems. For the $\Lambda$ scheme a very narrow linewidth is determined by the long relaxation rate of the ground state coherence. Counterpropagating lasers and one laser locked to an atomic transition are used to produce sub-Doppler resolution, as presented in~\cite{KrishnaNatarajan2005,WangHe2013,WangWang2014Cs} .\\
\indent{\it Frequency comb spectroscopy} (FC), The atomic absorption peaks are determined with a reference to a frequency comb. The absolute precision of the frequency readings is largely increased, see for instance~\cite{Udem1999,DasNatarajan2006b}. \\
\indent{\it Frequency modulated laser} (FML). When the exciting laser is modulated at the hyperfine splitting frequency, cross-over resonances are induced in three-level $\Lambda$ or V systems, as in~\cite{NobleWijngaarden2006}. The advantage of this technique is that only a single laser is required.\\ 
\indent{\it Hyperfine optical pumping and focus} (HOPF). Hyperfine transitions induced by microwaves or by optical pumping change the relative populations of the hyperfine levels of the ground state. In HOPF this modification is detected by measuring the atomic  beam intensity at the exit of a  magnet that focuses or defocuses atoms with different magnetic quantum numbers. The focused atoms are analyzed by a mass spectrometer, as in~\cite{Liberman1980}. \\
\indent {\it Hyperfine quantum beats} (HQB). Quantum beats are based on coherent pulsed excitation of excited hyperfine levels producing a time decay of the excited state populations modulated by the hyperfine frequency splitting. Polarized excitation and detection are required, as in~\cite{DeechSeries1977}. \cite{BelliniHaensch1997} applied delayed pulses of a frequency comb in order to probe the coherent hyperfine superposition of excited states. \\
\indent {\it Ion detection} (ION). This detection technique is very sensitive because a single ion can be detected.  It is applied within different schemes, such as the resonant laser ionization (RIS),  the selective electric field ionization of a Rydberg state or the thermoionic diode operation.\\ 
\indent {\it Laser-induced fluorescence spectroscopy} (LIF). The emitted fluorescence is monitored as a function of the laser frequency. In Doppler spectroscopy, a high resolution is  achieved  by a careful analysis of the absorption lineshapes, as in~\cite{TruongLuiten2015}.   In an AB with the laser propagation perpendicular to the atomic motion, the resolution is limited by the natural linewidth.  By applying a sudden change to the laser phase and monitoring the atomic evolution at a later time $T$, subnatural width resolution reaching  $\approx1/(2T)$ is achieved in~\cite{ShimizuTakuma1987}. For short-lived atoms with low density using a fast beam and a collinear laser propagation, LIF is  combined with nuclear decay to increase the signal to noise ratio as in~\cite{Duong1987,Lynch2016}. \\
\indent {\it  Level crossing by magnetic or electric fields}. An energy crossings of excited state levels vs an external parameter, either magnetic field (MLC) or electric field (ELC),  is monitored, see for instance~\cite{NagourneyHapper1978,Auzinsh2007}. A precise determination of  the applied field is required.\\
\indent {\it Maser} (MA). The emission frequency of a maser operating on a hyperfine ground state transition is measured in~\cite{Tetu1976}.\\
\indent{\it Magnetic field decoupling} (MFD). Starting from an initial anisotropic Zeeman sublevel population distribution, the hyperfine constants are derived from the polarization of the fluorescent emission monitored vs an applied magnetic field decoupling the nuclear and electronic angular momenta, as presented in~\cite{WijngaardenSagle1991b}.  \\
\indent{\it Microwave spectroscopy} (MWS). The population modifications induced by transitions between hyperfine levels, mainly in the microwaves,  are detected. In order to increase the signal-to-noise ratio, MWS is combined to other techniques, such as HOPF, LIF, or selective electric field ionisation  for Rydberg states as in~\cite{GoyHaroche1982}. The FOUNT+MWS combination applied to optical clocks leads to an extremely high precision, as in~\cite{GuenaBize2014,Ovchinnikov2015}. A Ramsey optical interferometer is used for the potassium ground state measurement of ~\cite{AriasWhitlock2019,PeperDeiglmayr2019}.\\ 
\indent{\it Optical radiofrequency or microwave double resonance} (ORFDR).  The radio-frequency induced transitions between excited states are detected through the modification of the LIF, either in its spectrum or in its polarization, as in~\cite{Farley1977,LamHapper1980}. Optical pumping is applied  to modify the population distribution and increase the detected signal.\\
\indent {\it Optical-optical double-resonance} (OODR). A two-colour excitation via an intermediate step produces the population of excited states.  The sub-Doppler resolution is obtained by operating  in a MOT in~\cite{FortTinoInguscio1995}, by applying the lasers in a counterpropagating geometry in~\cite{StalnakerTanner2010}, by a counterpropagating laser geometry selecting a single class of velocities different from zero in~\cite{LeeMoon2015}, or by using saturated absorption to lock on the first transition and excite only the atoms at zero velocity in~\cite{YangWang2016}.  Detection is based mainly on the spontaneous emission from the intermediate or final state. In presence of  an optical pumping process, the population distribution perturbed by the second step excitation is monitored, as in~\cite{WangWang2014Rb}.\\ 
\indent {\it Optical spectroscopy} (OS). Doppler limited high optical resolution spectroscopy from an alkali cell as in~\cite{TruongLuiten2015}, or Doppler-free in a MOT as in \cite{AntoniMicollierBouyer2017,AriasWhitlock2019}.\\ 
\indent {\it Resonant ionization spectroscopy} (RIS). Atoms in specific states are ionized by multistep laser absorption, and the ions are detected. It was introduced by~\cite{AndreevLethokov1987} for measuring of the ground state hyperfine structure in francium. The resonant tuning of the intermediate step provides the spectroscopic resolution of its hyperfine structure.    High resolution and sensitivity  due to ion detection, are associated with this technique as in the MOT experiment by~\cite{GabbaniniLucchesini1999} or for exotic isotopes in an atomic beam of exotic isotopes using 
the collinear laser spectroscopy as in~\cite{Lynch2014}. These last authors directed the ion  to an alpha-decay detection station for clear identification in order to reduce isobaric and ground state contamination in their francium isotopes studies. In order to increase the frequency resolution, in recent francium accelerated atomic beam experiments~\cite{Neugart2017}  the excitation laser is split into two beams, co-propagating and  counter-propagating with the atoms in order to increase the frequency resolution.\\
\indent{\it Saturated absorption spectroscopy} (SAS). This technique is based on a pump and probe laser applied to the same transition. It produces spectra with a natural linewidth resolution. The counter-propagating geometry compensates for the Doppler broadening. Main limitations are imposed by the laser stability, as analyzed in~\cite{DasNatarajan2008,GlaserFortagh2020} . \\
\indent {\it Stark spectroscopy} (SS). Stark spectroscopy is based on the electric field shift of atomic level energies. It is used mainly for Rydberg states as in~\cite{StevensMetcalf1995}. Information on lower energy states may be derived by the difference in level Stark shifts. \\ 
\indent {\it Two-photon sub-Doppler spectroscopy}  (TPSDS). A single-colour two photon not-resonant excitation explores highly excited states. The sub-Doppler resolution is obtained by operating with counterpropagating beams in a VC as described by~\cite{HerrmannWeis1985,HagelBiraben1999} or in a MOT as in~\cite{GeorgiadesKimble1994}. \\

\section{Hyperfine theory}

The hyperfine structure Hamiltonian $H_{hyp}$ of an atom having a single valence electron outside the closed shells consists of the magnetic dipole $H_{dip}$, the electric quadrupole $H_{quadr}$ and the octupole $H_{octup}$ terms \begin{equation}
H_{hyp} = H_{dip} + H_{quadr}+H_{octup}.
\end{equation}
\indent $H_{dip}$ describes the interaction of the nuclear magnetic moment with the  magnetic field generated by the electrons. For  the electron angular momentum $\vec{J}$ and  the nuclear angular momentum $\vec{I}$, it is given by
\begin{equation}
H_{dip} = hA\vec{I} \cdot \vec{J},
\end{equation}
where $A$ is the magnetic dipole constant and $h$ is the Planck constant.\\
\indent The electric quadrupole term originates from the Coulomb interaction between the electron and a nonspherically symmetric nucleus. It is given by
\begin{equation}
H_{quadr} = hB\frac{3(\vec{I}\cdot \vec{J})^2+\frac{3}{2}\vec{I}\cdot \vec{J}-I(I+1)J(J+1)}{2I(2I-1)J(2J-1)},
\end{equation}
where $B$ is the electric quadrupole moment coupling constant. This expression is valid for nuclear spins $I \ge 1$ and is zero otherwise.\\ 
\indent The octupole Hamiltonian $H_{octup}$ presented in~\cite{Armstrong1971} depends on the electron and nuclear tensor operators of rank 3, requires an electron angular momentum at least equal to 3/2 and is characterised by the hyperfine $C$ constant.  The explicit dependence on the atomic quantum numbers is detailed in \cite{GerginovTanner2003}.\\
\indent Hyperfine interactions decrease rapidly for higher-lying
states. ~\cite{Kopfermann1958} showed that the hyperfine constants are proportional to the expectation value of $1/r^{3}$, where $r$ is the distance between the nucleus
and valence electron. For highly excited electrons, the valence electron is far from the core electrons and $<1/r ^{3}>$  is well approximated by the hydrogenic result 
\begin{equation}
\label{1/r3}
<1/r ^{3}>\propto\frac{1}{(n^*)^3}\left(1-\frac{\partial \sigma}{\partial n}\right),
\end{equation}
where $n^*$ is the effective principal quantum number, $n$ is the principal quantum number, and the difference $\sigma(n)=n-n^*$ is the quantum defect. In a more refined treatment, by expressing the Schr\"odinger wavefunction at the nucleus position through the  effective nuclear charge $Z_I$ in the inner region where the orbit penetrates, and setting $Z_0=1$ for the net charge of the ion around which the single electron moves, the modified Fermi-Segre formula for the dipolar constant of the state with angular momentum  $l$ is derived in~\cite{Kopfermann1958}
\begin{equation}
A=\frac{8}{3}R_\infty\alpha^2g_I\frac{Z_iZ_0^2}{(n^*)^3}F_{r,J}(n, l ,Z )(1- \epsilon)(1- \delta)
\label{Aconstant}
\end{equation}
with $R_\infty$ being the Rydberg constant,  $\alpha$ the fine structure constant, and   $g_I$ the gyromagnetic ratio of the nuclear magnetic moment.  The relativistic  effects are expressed by the factor $F_{r,J}(n, l ,Z )$ near unity for light atoms and different from unity for large $Z$ numbers.  The $1- \epsilon$ factor is the change in the electronic wave function for distributions of  the nuclear charge over its volume. The $1- \delta$ factor is the change in the electron-nuclear  interaction by  the distribution  of  the magnetic moment, which is called the Breit-Weisskopf effect. \\
\indent For the quadrupole coupling constant $B$ the following expression is derived in~\cite{Kopfermann1958}: 
\begin{equation}
B=\frac{1}{h}\frac{e^2}{4\pi\epsilon_0}\frac{2J-1}{2J+2}Q<\frac{1}{r ^3}>R_{r,J}(n, l ,Z )
\label{Bconstant}
\end{equation}
with $e$ the charge of the electron, $\epsilon_0$ the vacuum permittivity, $Q$ the quadrupole nuclear moment and $R_{r,J}(n, l ,Z )$ a relativistic correction factor. From Eq.~\eqref{1/r3} for $1/r ^3$, it follows that the $B$ constant is also proportional to $1/(n^*)^3$.\\ 
\indent On the basis of the above expressions, where the nucleus is represented by a point charge, the following scaling  is derived in~\cite{Kopfermann1958} for the magnetic dipole and  magnetic quadrupole constants $A$ and $B$ associated with two different isotopes:
\begin{eqnarray}
\frac{A^i}{A^j}=\frac{g_I^i}{g_I^j}& \label{gIscaling},\\
\frac{B^i}{B^j}=\frac{Q^i}{Q^j}&\label{Qscaling},
\end{eqnarray} 
where $g_I^i$ and $Q^i$ represent the nuclear gyromagnetic factor $g$ and the quadrupole moment of the isotope $i$, respectively. Deviations from these isotopic scaling laws represent the hyperfine anomalies presented in Subsection \ref{Anomalies}.\\
\indent The dipolar and quadrupolar hfs Hamiltonians, involving the interaction of spin and orbital angular momenta with the nuclear moments,  have matrix elements diagonal over the hyperfine quantum numbers,  but also off-diagonal ones $(A_{J,J-1})$ connecting fine-structure states with different $J$ values as presented in~\cite{Arimondo1977}.  These  off-diagonal couplings produce a mixing of the eigenstates and a shift of the energies. For fine structure  states far apart in energy, a  perturbation of the hyperfine constants includes the influence of the off-diagonal matrix elements, as for the $6^2P$ Cs doublet in~\cite{JohnsonDerevianko2004}.  In the opposite case, such as the $2^2P$ $^7$Li doublet, the hyperfine splittings are expressed through the $a_i, (i=c,d,o)$ parameters of an effective hyperfine-splitting Hamiltonian: contact, dipolar and orbital, respectively, as derived in~\cite{LyonsDas1970}. In the first perturbation order and using the one-electron
theory, the different hyperfine constants are written as 
\begin{eqnarray}
A (^2P_{1/2}) &=& - a_c- 10a_d + 2a_o, \nonumber \\
A (^2P_{3/2}) &=& a_c + a_d+ a_o,\nonumber \\
A (^2P_{3/2},^2P_{1/2}) &=& -a_c +\frac{5}{4} a_d+\frac{1}{2} a_o.
\label{Offdiagonal}
\end{eqnarray}
These relations have been used by experimentalists for their data analysis.
Higher perturbation order corrections to the Li constants were derived in~\cite{BeloyDerevianko2008,PuchalskiPachucki2009}.  The mixing produced by off-diagonal  hyperfine interactions plays an important role in the cesium measurements for parity nonconservation, as in the experiments of ~\cite{GilbertWieman1986,BouchiatGuena1988}, and in the  theoretical analysis of~\cite{DzubaFlambaum2000}. It is expected to be even more important in francium.    \\
\indent Eq.~\eqref{Aconstant} predicts the following scaling law:
\begin{equation}
\label{Ascaling}
A\propto\frac{1}{(n^*)^3}.
\end{equation}
A similar one applies to $B$ on the basis of Eqs.~\eqref{1/r3} and~\eqref{Bconstant}.\\
\indent  For the $^{87}$Rb low-$n$ states, the $A$ dependence on $n^*$ was tested  in 1976 by~\cite{BelinSvanberg1976a} using the $1/(n^*)^2$ dependence of the fine structure data. The $A$ scaling law was later verified for $^{85}$Rb high $D$ states within 2 percent by~\cite{WijngaardenKoh1993}. More recently, with the very high-$n^*$ values having been precisely derived from laser spectroscopy, the scaling was tested for the $^{85}$Rb Rydberg states between $n=27$ and $n=33$   in~\cite{LiGallagher2003} and ~\cite{RamosRaithel2019}. The $^{87}$Rb data obtained by~\cite{LiGallagher2003} were reexamined for the scaling in~\cite{MackForthag2011}. In~\cite{Sassmannshausen2013} the scaling law was verified for the Cs  $^2S_{1/2}$ and $^2P_{1/2}$ Rydberg states in the $n=10-80$ range. \\ 
\indent The theoretical determination of the hyperfine constants has greatly evolved within the last few years. Instead of focusing on a few atomic states, the more recent calculations target a very large number of states.  While in 1999 a few  hyperfine constants  of different alkalis were calculated, e.g., in~\cite{SafronovaDerevianko1999},  more recently M. Safronova and coworkers in (~\cite{JohnsonSafronova2008,SafronovaSafronova2008,SafronovaSafronova2011,Auzinsh2007}) have produced a global derivation of the constants for $^7$Li,  $^{39}$K, $^{87}$Rb, and $^{133}$Cs. Later, the hyperfine data  for all the K isotopes were carefully examined by~\cite{SinghSahoo2012}, for Rb and Cs by ~\cite{GrunefeldGinges2019}, for Cs  by \cite{TangShi2019}, and for Fr  by \cite{SahooSakemi2015,LouTang2019,GrunefeldGinges2019}. The global analysis by~\cite{SinghSahoo2012} has derived precise values for the nuclear quadrupole moments of the potassium isotopes demonstrating a good internal consistency of the hyperfine data. The development led by Marianna Safronova to provide both experimental and theoretical energy level information for a large variety of atoms is a welcome addition to the data compilation. It is currently available online, see \cite{UDportal}.\\ 
\indent The hyperfine interaction and the weak interaction that gives rise to parity-nonconservation (PNC) in atoms both happen because the electron density overlaps with the nucleus. From a particle physics point of view, the exchange of the $Z^0$ boson carries the weak interaction with its PNC. Atomic PNC interest comes from its unique possibility to test the standard model at low energy. The structure of the nucleus is key to the details of nuclear-spin-independent PNC, where the electron axial-vector-current interacts with the nucleon vector-current. The nuclear-spin-dependent PNC, where the nucleon axial-vector-current interacts with the electron vector-current also depends on nuclear structure and is primarily due to the nuclear anapole moment.  As noted by~\cite{FlambaumKhriplovich1985} and confirmed by~\cite{BouchiatPiketty1991,JohnsonSafronova2003} the hyperfine interaction leads to the nuclear spin dependence of the matrix element in the atomic PNC.  Cs measurements by~\cite{WoodWieman1997} reached enough sensitivity to measure the anapole moment in the nuclear-spin-dependent part of the PNC interaction. A new generation of atomic parity violation experiments is underway, e.g. ~\cite{Gwinner2022}. These experiments are made with francium atoms. The PNC effects in Fr with respect to Cs are estimated 
to be 18 times larger for the nuclear-spin-independent and 11 times larger for the nuclear-spin-dependent part. \\
\indent The determination of the parity-conserving quantities in both high precision experiments and {\it{ab initio}} calculations, such as transition matrix elements, lifetimes, polarizabilities, and hyperfine constants, is essential for PNC studies. The hyperfine constants test in quantitative ways the quality of the electronic wavefunction near the nucleus.   This unique combination between theory and experiment has greatly favored the heaviest alkali atoms and has stimulated a large  search effort for the hyperfine structures in their isotopes as presented in~\cite{SafronovaClark2018}.

\section{Measured hyperfine constants}
\label{measuredconstants}
In the following, we present several Tables for the measured $A$ and $B$ constants of the alkali atoms. In a few cases, we derive  the $A$ constant from the measured hyperfine splitting reported by the authors using the formula for the hyperfine energies given by~\cite{Kopfermann1958}. In each Table, the atomic states are listed in order of increasing $n$, then of increasing $L$,  increasing $J$, and finally chronologically. Two columns report the acronyms determining  the atomic sample and the spectroscopic technique applied in the measurement. The reference to the original publication is in the last column.  The spectroscopic technique column reports the ``From'' notation for the data taken from~\cite{Arimondo1977}, for which a critical examination or a weighted averaging over several measurements was performed, leading to a recommended value.  When the hyperfine value reported in that reference remains the only one available, or its error bar is smaller than later measurements, the original work is directly quoted in the Table. Within the Table's $B$ column the entry "$0.$"  denotes that the authors have assumed the quadrupole constant equal to zero.\\
\indent A few techniques, such as the HQB, are not able to resolve the sign of the constants for the explored state. On the basis of the above scaling laws applied to different atoms within Section \ref{Scalinglaws}, we have produced a dependable sign assignment for most states. If the sign was not identified, the absolute value is reported.\\   
\indent  The measurement uncertainties are reported in the text and Tables in parentheses after the value, in units of the last decimal place of the value. For example, 153.3(11) means 153.3 $\pm$1.1. Most authors specify their uncertainty on the level of one standard deviation. If a different convention is used, it is mentioned in the text.\\
\indent For each atomic species, we  mention in the text the states where a very high precision is obtained or where a disagreement between the measured values exists. When several ($n$) measurements  are associated with a single state,  the Tables include a weighted average, w.a., representing a reference for further work. We follow the procedure of the Particle Data Group in \cite{PDG} in the Introduction, Sec. 5.2.2,  {\it Unconstrained averaging}, to find the weighted error (w.e.). We calculate it first based on the $n$ individual errors ($e_i$), w.e.$={(1/\sum {1/e_i}^2)^{1/2}}$ 
We also calculate the reduced $\chi$-squared $(\chi^2_{\rm{red}})$ with $n-1$ degrees of freedom, to test the size of the weighted error. If $(\chi^2_{\rm{red}})$ is greater than unity by more than one standard deviation  $(2/(n-1))^{1/2}$, then we increase the w.e.  of the w.a. by the factor $(\chi^2_{\rm{red}})^{1/2}$,  so that the weighted enhanced error (w.e.e.) is w.e.e.= $(\chi^2_{\rm{red}})^{1/2}\times$w.e..
We report in the table either the w.a. with its weighted error or the enhanced error (w.e.e.), which we explicitly state.
Such averaging is not performed when  the precision of one measurement is greater than all the remaining ones, which is then denoted by ``Recommended'' in the Table's last column. The last column contains a ``See text'' statement, if one or more values are not included into the w.a..\\  
\indent A different way of averaging developed by \cite{Rukhin2009,Rukhin2019} evaluates the clustering of the data and assigns individual hidden uncertainties to the measurements from different groups. These uncertainties are then added in quadrature to the stated uncertainties.  Weighted averages and weighted errors calculated by this cluster maximum likelihood estimator (CMLE) are not always identical to the values reported in our Tables.  For completeness, whenever a Table recommended value includes a w.e.e. derived from the $(\chi^2_{\rm{red}})^{1/2}$  analysis, we have used the CMLE method to calculate the corresponding w.a.$_{\rm CMLE}$ weighted average and w.e.e.$_{\rm CMLE}$ weighted enhanced error. The comparison between the recommended values obtained by these two approaches  shows that in almost all the
28 cases the weighted average of the two analyses
agrees within two times the w.e.e.. There are only two cases with a greater difference, associated with a large $(\chi^2_{\rm{red}})^{1/2}$ value as an indicator of an anomalous scatter of the data. The interested reader can find results and plots  in the  Supplementary Information (SI) of this review, reference~\cite{SI_Hyperfine}.

\subsection{ Lithium} Data for this atom are reported in Table~\ref{Table:Li}. As for all alkalis, several spectroscopic investigations are stimulated by the interest in laser cooling, but  only a few experiments are performed in a MOT.  The recent $^6$Li MOT experiments by~\cite{WuWu2018,LiWu2020,RuiWu2021} at Shanghai may open a new trend.  For the ground state of both lithium isotopes, the "old'' atomic beam measurements based on MWS report the highest precision, not  reached by the several ones based on laser spectroscopy. \cite{OttoHuettner2002} performed measurements for a few high-$n$ states of $^7$Li, as well as for other alkalis.\\
\indent  For the $^{6,7}$Li 2$^2P_{1/2,3/2}$ state measurements by~\cite{Umfer1992}, the error bar is derived on the basis of Eqs.~\eqref{Offdiagonal} from  the off-diagonal constants presented in the following.  For the $A$ value of the 3$^2S_{1/2}$ state of both isotopes measured by~\cite{LienLiu2011}, we estimated the error bar not reported by the authors. \\
  \indent A wide experimental effort concentrates on the $2^{2}P_{1/2}$ state of both isotopes. Instead, the $2^{2}P_{3/2}$ state of $^6$Li remains with the data obtained before laser spectroscopy. For $^7$Li, three experimental investigations of this doublet, by~\cite{Orth1975}, ~\cite{NagourneyHapper1978} and ~\cite{Umfer1992},
consider the contribution of the off-diagonal hyperfine couplings linked to the small fine-structure splitting. \cite{Orth1975} include in their analysis the previous level crossing data by~\cite{BrogWieder1967}. The  analysis by ~\cite{NagourneyHapper1978}  combines their own data and  the previous ones. ~\cite{Umfer1992} acquired enough data for their own analysis. ~\cite{Orth1975} using Eqs.~\eqref{Offdiagonal} and their measured $A_{3/2,1/2}=11.85(35)$ MHz found $a_c=-9.838(48)$, $a_d=- 1.876(12)$, $a_o=8.659(37)$. ~\cite{NagourneyHapper1978} reported $a_c=-9.838(48)$, $a_d=- 1.975(22)$, $a_o=8.659(37)$,  and ~\cite{Umfer1992} obtained $a_c=- 9.93(37)$, $a_d=-1.72(20)$, $a_o=8.69(31)$ (all in MHz).  For the same  doublet,~\cite{BeloyDerevianko2008} performed an accurate evaluation of the off-diagonal hyperfine couplings. They produced corrections to the measured dipole constant in the $A(^2P_{1/2})$ accurately measured by ~\cite{Orth1975}, ~\cite{WallsWijngaarden2003}, and ~\cite{DasNatarajan2008}. For this last one, the 27.0 kHz correction  should be compared to the author's original 3 kHz experimental uncertainty.\\
\indent A disagreement exceeding the error bars exists for the 3$^2S_{1/2}$ $^7$Li hyperfine constant $A$ measured  in~\cite{StevensMetcalf1995}, compared to those of~\cite{BushawDrake2003,LienLiu2011}. We do not include that measurement in the w.a.\\
\indent The discrepancy for the $3^2P_{3/2}$ $^7$Li constants between the previous values by\cite{BudickNovick1966,IslerNovick1969}  and the ~\cite{NagourneyHapper1978} ones remains unexplained as pointed out by these authors. Therefore, instead of the recommended value from \cite{Arimondo1977}, we consider all the previous measurements.   The $\chi^2_{red}$ correction is applied to the statistical error. An analysis of the off-diagonal elements was performed by~\cite{NagourneyHapper1978} for the $3^2P$ $^7$Li doublet, leading to $a_c=-3.10(67)$, $a_d=- 0.54(27)$, $a_o=2.61(40)$.\\ 
\begin{longtable*}{cddccl}
\caption{Measured $A$ and $B$ values for Li isotopes\label{Table:Li}}\\
\hline
State& A (MHz) & B(MHz)&Sample&Technique  &Ref.  \\
\hline
\endfirsthead
\multicolumn{6}{c}%
{\tablename\ \thetable\ -- \textit{Continued from previous page}} \\
\hline
State& A (MHz) & B(MHz)&Sample&Technique  &Ref.   \\
\hline
\endhead
\hline \multicolumn{6}{r}{\textit{Continued on next page}} \\
\endfoot
\hline
\endlastfoot
&&&{\bf $^{6}$Li} && \\
2$^2S_{1/2}$&152.1368407(20)&   -  &AB&MWS&From~\cite{Arimondo1977} \\
"                    &151.(3)                     &   -  &VC&ION&~\cite{LorenzenNiemax1982} \\
"                    &153.3(11)                 &   -  &AB&LIF&~\cite{Windholz1990} \\
"                    &152.109(43)             &   -  &AB&LIF&~\cite{WallsWijngaarden2003} \\
"                    &152.121(57)             &    -  &AB&FML&~\cite{NobleWijngaarden2006}   \\
"                    &152.143(11)              &   -  &AB&LIF+FC&~\cite{SansonettiPorto2011}\\ 
"                    &152.1343(9)              &   -  &MOT&LIF&~\cite{WuWu2018,LiWu2020} \\
"                    &152.1368407(20)      &   -  &-&-&Recommended\\
2$^2P_{1/2}$&17.375(18)    &   -             &VC&ORFDR&~\cite{OrthOtten1974}\\
"                    &17.8(3)           &   -            &VC&MLC&~\cite{NagourneyHapper1978}\\
"                    &16.81(70)       &   -            &AB&LIF&~\cite{Windholz1990} \\
"                    &17.386(31)     &   -            &AB&LIF&~\cite{WallsWijngaarden2003} \\
"                    &17.407(37)     &   -            &AB&FML&~\cite{NobleWijngaarden2006} \\
"                    &17.394(4)       &   -             &AB&LIF      &~\cite{DasNatarajan2008} \\
"                    &17.407(10)     &   -             &AB&LIF+FC&~\cite{SansonettiPorto2011}\\
"                    &17.4021(9)     &   -             &MOT&LIF&~\cite{LiWu2020,RuiWu2021}\\
"                    &17.408(13)     &   -             &MOT&LIF+DD&~\cite{LiWu2021}\\
"                    &17.4017(9)      &   -            &-&-&~w.a.\\
2$^2P_{3/2}$&-1.155(8)        & -0.10(14)  &AB&ORFDR&~\cite{OrthOtten1974}\\
3$^2S_{1/2}$&34.(13)            &   -             &VC&ION&~\cite{VadlaNiemax1987} \\
"                    &35.263(15)      &   -             &AB&TPSDS+RIS&~\cite{BushawDrake2003} \\
"                    &35.283(10)      &   -             &AB&RIS&~\cite{EwaldZimmermann2004} \\
"                    &35.20(20)        &   -             &AB&LIF&~\cite{LienLiu2011}\\
"                    &35.267(14)      &   -             &AB&RIS&~\cite{Nortershauer2011}\\
"                    &35.274(7)&      -                   &-&-&~w.a.\\
3$^2P_{1/2}$&5.3(4)             & -               &VC&MLC&~\cite{NagourneyHapper1978}\\
3$^2P_{3/2}$&-0.40(2)          &0.              &VC&MLC&~\cite{IslerNovick1969} \\
4$^2S_{1/2}$&13.1(13)         &-                &AB  &LIF&~\cite{KowalskizuPutlitz1978}\\
"                    &15.(3)              &-                &VC&ION&~\cite{LorenzenNiemax1982} \\
"                    &13.5(8)            &-                 &AB  &LIF&~\cite{DeGraffenreidSnsonetti2003}\\
"                    &13.5(7)            &   -             &-&-&~w.a.\\
\hline
 &&&{\bf $^{7}$Li}&&  \\
  2$^2S_{1/2}$&401.7520433(5) &   -  &AB&MWS&~\cite{BeckmannElke1974} \\
  "                    &401.(5)                &   -      &VC&ION&~\cite{LorenzenNiemax1982} \\
  "                    &401.81(25)          &   -  &AB&LIF&~\cite{Windholz1990} \\
  "                    & 401.767(39)       &   -     &AB&LIF&~\cite{WallsWijngaarden2003}\\
  "                    & 401.772(33)       &   -     &AB&MFL&~\cite{NobleWijngaarden2006}\\
  "                    & 401.747(7)         &-  &AB&LIF+FCS&~\cite{SansonettiPorto2011} \\
   "                   & 401.755(8)         &   -     &AB&LIF&~\cite{HuangWang2013} \\
  "                    &401.7520433(5)  &   -     &-&-&Recommended \\
  2$^2P_{1/2}$&45.914(25)          &-        &  AB  &ORFDR&~\cite{OrthOtten1974}\\
  "                    & 46.05(30)            &   -     &AB&LIF&~\cite{Windholz1990} \\
  "                    & 46.175(2980)      &-          &AB&MLC&~\cite{Umfer1992} \\
  "                    & 46.010(25)          &   -  &AB&LIF&~\cite{WallsWijngaarden2003}\\
  "                    & 45.893(26)          &   -  &AB&FML&~\cite{NobleWijngaarden2006} \\
  "                    & 46.024(3)            &   -  &AB&LIF&~\cite{DasNatarajan2008} \\
  "                    & 46.047(3)            &   -  &VC&SAS&~\cite{SinghNatarajan2010} \\
  "                    & 45.938(5)             &   -  &AB&LIF+FCS&~\cite{SansonettiPorto2011} \\
  "                    & 45.946(4)            &   -  &AB&LIF&~\cite{HuangWang2013} \\
  "                    & 46.005(16)            &   -     &-&-&~w.a.(w.e.e.)\\
2$^2P_{3/2}$ &-3.055(14)              &-0.221(29)  &AB&ORFDR&~\cite{Orth1975}\\
"                      & -2.95(4)                &0.           &VC &MLC&~\cite{NagourneyHapper1978}\\
"                      & -3.08(4)                &-0.16(10) &AB &LIF+DD&~\cite{ShimizuTakuma1987} \\
"                      &-3.18(10)               &-0.8(7)             &AB&LIF&~\cite{Windholz1990}\\
"                      & -3.08(8)                & -0.20(27)  &AB &HQB&~\cite{CarlssonSturesson1989} \\
"                      &-2.96(88)               &-0.1(5)          &AB&MLC&~\cite{Umfer1992}\\
"                      & -3.050(16)            & -0.22(3)     &-&-&~w.a.\\
3$^2S_{1/2}$  &95.(10)                    &   -             &VC&ION&~\cite{VadlaNiemax1987} \\
"                      &94.68(22)                &    -             &AB&SS&~\cite{StevensMetcalf1995} \\
"                      &93.106(11)               &   -             &AB&TPSDS+RIS&~\cite{BushawDrake2003} \\ 
"                      &93.117(25)               &   -             &AB&RIS&~\cite{EwaldZimmermann2004} \\
 "                     &93.13(20)                 &   -             &AB&LIF&~\cite{LienLiu2011} .\\
 "                     &93.103(11)               &   -             &AB&RIS&~\cite{Nortershauer2011}\\
 "                     &93.095(52)               &   -             &AB&SAS&~\cite{KumarNatarajan2017} \\
 "                     &93.105(7)               & -                 &-&-& ~See text. w.a. \\
 3$^2P_{1/2}$&13.5(2)                &-                      &VC&MLC&~\cite{BudickNovick1966}\\ 
 "                    &13.7(12)              &-                 &VC&MLC&~\cite{NagourneyHapper1978}\\         
"                     &13.5(2)                & -    &-&-&Recommended \\
3$^2P_{3/2}$&-0.96(13)             &0. &VC&MLC&\cite{BudickNovick1966}\\
"                    &-0.965(20)           &-0.019(22) &VC&MLC&\cite{IslerNovick1969}\\
"                    &-1.036(16)           &-0.094(10) &VC &MLC&~\cite{NagourneyHapper1978} \\
"                    &-1.01(2)               & -0.081(28)  &-&-&~w.a. (w.e.e.) \\
3$^2D_{3/2}$& 0.843(41)            & 0.               &AB&TPSDS&~\cite{Burghardt1988} \\
"                     & 1.14(49)              & 0.               &VC&TPSDS    &~\cite{OttoHuettner2002} \\
"                     &0.843(41)            & 0.               &-&-&Recommended\\
3$^2D_{5/2}$ & 0.3436(10)          &0.                &AB &TPSDS&~\cite{Burghardt1988} \\
"                      & 0.31(13)              &0.                &VC &TPSDS&~\cite{OttoHuettner2002} \\
"                      &0.3426(10)           &0.                &-&-&Recommended \\
4$^2S_{1/2}$  & 36.4(40)               &-               &AB  &LIF&~\cite{KowalskizuPutlitz1978}\\
"                       &38.(3)                    &   -            &VC&ION&~\cite{LorenzenNiemax1982} \\
"                       & 35.32(72)             &-               &VC &TPSDS&~\cite{OttoHuettner2002} \\
"                       & 34.9(4)                 &-               &AB  &LIF&~\cite{DeGraffenreidSnsonetti2003}\\
"                       &35.05(35)             & -               &-&-&~w.a. \\
4$^2P_{3/2}$  &-0.41(4)                 & 0.            &VC&MLC&~\cite{IslerNovick1969}\\
6$^2S_{1/2}$  &38.0(15)                 & -              &VC  &TPSDS&~\cite{OttoHuettner2002} \\
7$^2S_{1/2}$  & 9.2(25)                 &  -             &" &"&"     
\end{longtable*}

\subsection{Sodium}
The sodium results of Table~\ref{Table:Na} are emblematic of the progress achieved in hyperfine constant measurements. In chronological order, the two-photon sub-Doppler spectroscopy was applied in 1978 to probe the $4^2D$ excited states, by~\cite{BirabenBeroff1978} and ~\cite{Burghardt1978}, extended by this last research group to the $3^2D$ states in~\cite{Burghardt1988}. In 1989, ~\cite{KasevichChuDeVoe1989} published the first hyperfine ground state determination in an atomic fountain with 1 mHz precision, close to the previous best atomic beam value of Table~\ref{Table:Na}, opening the road to further amazing improvements in fountain atomic clocks. ~\cite{ZhuHall1993}  performed the first hyperfine constant measurement in a MOT exploring the $5^2P$ states with a relative precision $\approx1\times10^{-4}$ among the best ones for excited states. In the same year,  ~\cite{YeiSieradzan1993} performed subnatural linewidth measurements of hyperfine coupling constants using the delayed detection in polarization quantum beat spectroscopy. In 2003, the excitation of ultracold atoms in a MOT on the $3^2P-4^2P$ electric quadrupole transition allowed ~\cite{BhattacharyaBigelow2003}  to perform high resolution spectroscopy of the 4$^2P_{1/2}$ level. ~\cite{DasNatarajan2006} introduced the CCS approach for the first excited state, 3$^2P_{3/2}$. However, no competitive new data  for the ground state are available. \\
\indent The 3$^2P_{1/2,3/2}$ states have been examined by several authors, with an increasing precision and a good agreement among the different results. 
For the 4$^2D$ states of~\cite{BirabenBeroff1978} and for the 7, 8 and 9$^2P_{3/2}$
ones of~\cite{JiangLundberg1982}, the sign has been determined here from the scaling law. In sodium, there is not enough data for a global analysis. For the large majority of excited states above the $5^2P$ ones, no new data have been published after 1982. 
\begin{longtable*}{cddcclp{2.0cm}}
\caption{Measured $A$ and $B$ values for $^{23}$Na\label{Table:Na}}\\
\hline
State& A (MHz) & B(MHz)&Sample&Technique  &Ref.   \\
\hline
\endfirsthead
\multicolumn{6}{c}
{\tablename\ \thetable\ -- \textit{Continued from previous page}} \\
\hline
State& A (MHz) & B(MHz)&Sample&Technique  &Ref.   \\
\hline
\endhead
\hline \multicolumn{6}{r}{\textit{Continued on next page}} \\
\endfoot
\hline
\endlastfoot
3$^2S_{1/2}$&885.8130644(5) &  -          &AB&MWS& From~\cite{Arimondo1977} \\
"                    &885.70(25)         &  -          &VC&SAS&~\cite{Pescht1977} \\
"                    &885.813065(1)   &  -          &FOUNT&RIS&~\cite{KasevichChuDeVoe1989} \\
"                    &885.8130644(5)  &  -         &-&-&~Recommended \\
3$^2P_{1/2}$&94.25(15)            &  -          &VC&SAS&~\cite{Pescht1977}\\
"                    &94.47(1)              &  -          &AB&LIF&~\cite{GriffithvanZyl1977} \\
"                    &94.05(20)            &  -          &AB&LIF&~\cite{Umfer1992} \\
"                    &94.42(19)            &  -          &AB&HQB&~\cite{CarlssonFroeseFisher1992} \\
"                    &94.44(13)            &  -          &AB&LIF&~\cite{vanWijngaarden1994} \\ 
"                    &94.7(2)                &-             &AB&LIF&~\cite{ScherfWindholz1996} \\
"                    &94.349(7)            & -             &VC&SAS&~\cite{DasNatarajan2008} \\
"                    &94.39(2)              &  -            &-&-&~w.a.(w.e.e.) \\
3$^2P_{3/2}$&18.64(6)             &2.77(6)          &AB&HQB&~\cite{KristAndrae1977} \\
"                    &18.69(6)              &2.83(10)        &AB&HQB&~\cite{CarlssonSturesson1989} \\
"                    &18.78(8)              &2.60(41)        &AB&LIF&~\cite{Umfer1992} \\
"                    &18.534(15)          &2.724(30)      &VC&HQB+DD&~\cite{YeiSieradzan1993} \\
"                    &18.62(21)            &2.11(52)        &AB&LIF&~\cite{vanWijngaarden1994} \\
"                    &18.8(1)                &2.7(2)            &AB&LIF&~\cite{ScherfWindholz1996} \\
"                    &18.79(12)            &2.75(12)         &AB&HQB&~\cite{VolzSchmoranzer1996} \\
"                    &18.572(24)          &2.723(55)       &AB&LIF&~\cite{Gangrsky1998} \\
"                    &18.530(3)            &2.721(8)          &VC&CCS&~\cite{DasNatarajan2006Na} \\ 
"                    &18.532(6)            &2.722(7)          &-&-&~w.a.(w.e.e.); w.a.\\
3$^2D_{3/2}$ &0.527(25)           &0.                   &AB&TPSDS&~\cite{Burghardt1988} \\
3$^2D_{5/2}$ &0.1085(24)         &0.                   &"&"&~" \\
4$^2S_{1/2}$ &203.6(2)             &-                     &VC&TPSDS+LIF&~\cite{Arqueros1988}  \\
4$^2P_{1/2}$ & 30.4(5)              &-                     &VC&ORFDR&~\cite{GrundevikLundberg1978} \\
"                     & 30.6(1)              &-                     &MOT&OODR+RIS&~\cite{BhattacharyaBigelow2003} \\
"                     &30.6(1)                &-                     &-&-&~w.a.\\
4$^2P_{3/2}$ & 6.022(61)          &0.97(6)           &VC&LC+ORFDR&From~\cite{Arimondo1977} \\
4$^2D_{3/2}$ &0.23(12)             &0.                   &VC&TPSDS&~\cite{BirabenBeroff1978}   \\
"                      &0.215(15)           &0.                   &AB&TPSDS&~\cite{Burghardt1978}   \\
"                      &0.215(15)           &0.                   &-&-&~Recommended\\
4$^2D_{5/2}$&<|0.28|               &0.                   &VC&TPSDS&~\cite{BirabenBeroff1978} \\
"                      &0.029(6)              &0.                  &AB&TPSDS&~\cite{Burghardt1978}\\
"                      &0.029(6)              &0.                  &-&-&Recommended  \\
5$^2S_{1/2}$ &77.6(2)                 & -                   &VC&ORFDR&~\cite{{TsekerisGupta1976}} \\
"                      &77.2(2)                 & -                  &MOT&TPSDS+RIS&~\cite{MarcassaBagnato1998} \\
"                      &77.40(14)                 &-                   &-&-&~w.a.\\ 
5$^2P_{1/2}$ &13.3(2)                  &-                   &VC&ORFDR&~\cite{GrundevikLundberg1978} \\
"                      &13.4687(42)         &-                   &MOT&SAS&~\cite{ZhuHall1993} \\
"                      &13.4687(42)         &-                   &-&-&~Recommended\\
5$^2P_{3/2}$ &2.64(1)                 &0.38(3)          &AB&HQB&~\cite{GrundevikSvanberg1979} \\ 
"                      &2.6360(23)          &0.3704(77)   &MOT&SAS&~\cite{ZhuHall1993} \\
"                      &2.6360(23)          &0.3704(77)     &-&-&~Recommended\\
6$^2S_{1/2}$  &37.5(2)                &-                     &VC&ORFDR&~\cite{LundbergSvanberg1977} \\ 
"                      &34.5(45)              &-                      &AB&SS&~\cite{HawkinsSchawlowSvanberg1977}\\
"                      &37.5(2)                &-                      &-&-&Recommended\\ 
6$^2P_{3/2}$  &1.39(1)                &  0.21(2)          &AB&HQB&~\cite{GrundevikSvanberg1979} \\
7$^2S_{1/2}$  &20.9(1)                &  -                     &AB&ORFDR&~\cite{LundbergSvanberg1977}\\
"                      &23.3(65)               &  -                    &AB&SS&~\cite{HawkinsSchawlowSvanberg1977}\\
"                      &20.9(1)                 &-                       &-&-&~Recommended\\ 
7$^2P_{3/2}$ &0.82(1)                   &0.13(3)  &VC&QBS&~\cite{JiangLundberg1982}\\
8$^2S_{1/2}$ &12.85(10)               &- &VC&ORFDR&~\cite{LundbergSvanberg1977}\\
8$^2P_{3/2}$&0.535(15)                &0.070(25)  &VC&QBS&~\cite{JiangLundberg1982}\\
9$^2P_{3/2}$&0.36(1)                    &0.045(15)  &"&"&~"\\
\end{longtable*}

\subsection{Potassium}
Our previous Li and Na remarks on the ground state values do not apply to the $^{39}$K ground state, as shown in Table ~\ref{Table:K}. For this atom, several recent measurements exist, with the frequency comb spectral resolution applied to determine the absolute transition frequencies, and from them, the fine and hyperfine splittings. A high precision is achieved, usually with a good agreement among data from different research groups, e. g.  for several $n=5-8$ $^2S_{1/2}$ states with measurements performed over a long time span. ~\cite{OttoHuettner2002} explored the $^2S$ states of the 39 and 41 isotopes  up to $n=14$   using two-photon sub-Doppler spectroscopy. \\ 
\indent  The  spatial dependence of  the HQB polarized fluorescence intensity in a given magnetic field was used by~\cite{GlodzKrainska1985} to derive for the first time the signs of the $A$ constants in the $6^2D$ states of $^{39}$K. For other $^2D$ states, the investigations by~\cite{BelinSvanberg1975}, ~\cite{Sieradzan1997}and ~\cite{GlodzKrainska1985b}  produced only the absolute sign of the $A$ and $B$ constants. Their signs are determined here on the basis of the scaling laws.  
\subsubsection{$^{39}K$}
The ground 4$^2S_{1/2}$ state was measured in three  MOT experiments by~\cite{AntoniMicollierBouyer2017,AriasWhitlock2019,PeperDeiglmayr2019}. ~\cite{PeperDeiglmayr2019} obtained a value close to the  old atomic beam experiments with a difference in the 10 Hz range.  ~\cite{AriasWhitlock2019} claimed that their small discrepancy with the previous AB measurement can be accounted for by the $16(6)$ Hz quadratic Zeeman shift of the bias field and by the differential ac Stark shift in the optical dipole trap. The Table~\ref{Table:K} value takes into account such shift.~\cite{AntoniMicollierBouyer2017}  reported an all-optical measurement of the hyperfine splitting with a low statistical uncertainty, but there were uncontrolled systematical errors in their work, according to ~\cite{PeperDeiglmayr2019}.   \\
\indent A large disagreement exists between the measured $A$ values of the  4$^2P_{1/2}$ states, some of them having been reported with high precision.  The data are centred around two separate values 27.78(4) and 28.849(5) MHz with a separation larger than the reported precision. The lower values are by~\cite{BendaliVialle1981,TouchardVialle1982K,Duong1982,PapugaYordanov2014}, (all of them with low precision)  and by~\cite{FalkeGrosche2006} with  higher precision. The greater values were obtained in an AB magnetic resonance experiment by~\cite{BuckRabi1957} with a low precision and in two recent measurements by the Bangalore research group~\cite{BanerjeeNatarajan2004,DasNatarajan2008}  used the accurate CCS technique. The $\chi^2_{\rm{red}}$ for the full data set leads to a very large error bar.  Light shifts corrections, an important issue for several recent publications, were taken into account by the Bangalore group.  They stated,  "we do not have a satisfactory explanation for such a large discrepancy''. Such a discrepancy does not exist for the 4$^2P_{3/2}$ values determined at the same time by~\cite{FalkeGrosche2006} and~\cite{DasNatarajan2008}. ~\cite{FalkeGrosche2006} compared their  $^{6,7}$Li $D$  optical transition values to those by~\cite{BanerjeeNatarajan2004} and attributed the discrepancies to systematic errors in the laser calibration, more precisely to phase shifts in the wavelength (not frequency) comparison of the atomic excitation lasers. In~\cite{DasNatarajan2008}, the  driving frequency of an acousto-optical modulator gives a direct measurement of the hyperfine interval and the calibration issue should have been resolved. For the discrepancies in those optical frequencies,  \cite{BrownGillaspy2013} pointed out the important role of quantum interference and light polarization effects.  For the 4$^2P_{1/2}$ state the Table \ref{Table:K} reports a w.a. excluding the ~\cite{BanerjeeNatarajan2004,DasNatarajan2008} values.  \\
\indent For the 6$^2S_{1/2}$ state, the $A$ value referred to~\cite{ThompsonStoicheff1983} in the Table \ref{Table:K} was derived from their hyperfine splitting in~\cite{StalnakerRowan2017}.\\
\subsubsection{$^{41}K$} The 7$^2S_{1/2}$ values by~\cite{ThompsonStoicheff1983} and by~\cite{OttoHuettner2002} have a large disagreement. The first value is not consistent with the $1/(n^*)^3$ scaling law applied to the $n^2S_{1/2}$ states of this isotope. The ~\cite{OttoHuettner2002} value  is the ``Recommended'' one.\\
\begin{longtable*}{cddccl}
\caption{Measured $A$ and $B$ values for K isotopes\label{Table:K}}\\
\hline
State& A (MHz) & B(MHz)&Sample&Technique  &Ref.  \\
\hline
\endfirsthead
\multicolumn{6}{c}
{\tablename\ \thetable\ -- \textit{Continued from previous page}} \\
\hline
State& A (MHz) & B(MHz)&Sample&Technique  &Ref.   \\
\hline
\endhead
\hline \multicolumn{6}{r}{\textit{Continued on next page}} \\
\endfoot
\hline
\endlastfoot
&&&{\bf $^{39}$K}&&\\  
3$^2D_{3/2}$&  <|1.8|           &0.           &VC&ORFDR&~\cite{LamHapper1980}\\
"                    &0.96(4)              &0.37(8)  &VC&HQB+DD&~\cite{Sieradzan1997}\\
"                    &0.96(4)              &0.37(8)  &-&-&~Recommended\\
3$^2D_{5/2}$& <|2.2|             &0.              &VC&ORFDR&~\cite{LamHapper1980} \\
"                     &-0.62(4)             &<|0.3|      &VC&HQB+DD&~\cite{Sieradzan1997} \\
"                     &-0.62(4)            &0.             &-&-&~Recommended\\
4$^2S_{1/2}$ &230.8598601(3)&   -              &AB&MWS&From~\cite{Arimondo1977} \\  
"                     &231.0(3)             &-                 &AB&LIF   &~\cite{TouchardVialle1982K}\\
"                     & 230.8599(1)      &-                 &AB&MWS   &~\cite{Duong1993}\\
"                     &213.0(3)             &-                 &AB&LIF   &~\cite{PapugaYordanov2014}\\
"                     &230.859858(6)   &-           &MOT&MWS   &~\cite{AriasWhitlock2019}\\
"                     &230.859850(3)    &-           &MOT&MWS   &~\cite{PeperDeiglmayr2019}\\
"                     &231.1(3)               &-                      &AB&RIS   &~\cite{KoszorusWilkins2019}\\
 "                    &230.8598601(3)   &-           &-&-   &~Recommended\\
4$^2P_{1/2}$ &28.85(30)             &-                  &AB&MWS    &~\cite{BuckRabi1957}\\
"                     &27.80(15)              &-                  &AB&FML    &~\cite{BendaliVialle1981,Duong1982}\\
"                     &27.5(4)                  &-                  &AB&LIF    &~\cite{TouchardVialle1982K}\\
"                     &28.859(15)            &-                  &VC&SAS&~\cite{BanerjeeNatarajan2004}\\
"                     &27.775(42)             &-                  &AB&LIF+FC&~\cite{FalkeGrosche2006}\\              
"                     &28.848(5)               &-                  &VC&CCS&~\cite{DasNatarajan2008}\\
"                     &27.8 (2)                  &-                  &AB&LIF   &~\cite{PapugaYordanov2014}\\
"                     &27.793(71)           &-                &-&-   &~See Text. w.a.(w.e.e.)  \\
4$^2P_{3/2}$&6.093(25)          &2.786(71)   &AB&LIF+FC&~\cite{FalkeGrosche2006}\\
"                    &6.077(23)          &2.875(55)   &VC&CCS&~\cite{DasNatarajan2008}\\
"                    &6.084(17)          &2.842(43)    &-&-   &~w.a. \\
5$^2S_{1/2}$&55.50(60)          &-                  &VC&ORFDR&~\cite{GuptaSvanberg1973}\\
5$^2P_{1/2}$&9.02(17)            &-                  &VC&ORFDR&From \cite{Arimondo1977}\\
"                    &8.93(69)            &-                  &VC&SAS&\cite{HalloranBehr2009}\\
"                    &9.01(17)            &-                  &-&-   &~w.a. \\
5$^2P_{3/2}$&1.969(13)          &0.870(22)    &VC&MLC&From~\cite{Arimondo1977}\ \\
5$^2D_{3/2}$&0.44(10)            &0.                 &VC&ORFDR&\cite{BelinSvanberg1975} \\
5$^2D_{5/2}$&-0.24(7)            &0.                 &"&"&" \\
6$^2S_{1/2}$&21.81(18)          &-                    &VC&ORFDR&~\cite{GuptaSvanberg1973}\\
"                    &20.4(23)            &-                    &TD&TPSDS&~\cite{ThompsonStoicheff1983} \\
"                    & 21.8(5)             &-                    &VC&TPSDS&~\cite{KiranKumar2011} \\
"                    & 21.93(11)         &-                    &VC&OODR+FC&~\cite{StalnakerRowan2017} \\
"                    &21.89(9)            &-                   &-&-   &~w.a. \\
6$^2P_{1/2}$&4.05(7)             &-                    &VC&ORFDR&\cite{BelinSvanberg1975} \\
6$^2P_{3/2}$&0.886(8)           &0.370(15)     &VC&MLC&From~\cite{Arimondo1977}\\
6$^2D_{3/2}$& 0.25(1)            &0.05(2)         &VC&HQB+MFD&~\cite{GlodzKrainska1985} \\
6$^2D_{5/2}$& -0.12(4)           &0.                 &"&"&"\\
7$^2S_{1/2}$&10.79(5)            &-                   & VC&ORFDR+MLC&From~\cite{Arimondo1977}\\
"                    &12.7(24)            &-                   &TD&TPSDS&~\cite{ThompsonStoicheff1983} \\
"                    &10.41(93)         &-                    &VC&TPSDS&~\cite{OttoHuettner2002} \\
"                    &10.79(5)            &-                   &-&-   &~w.a. \\
7$^2P_{1/2}$  &2.18(5)            &-                   &VC&ORFDR&\cite{BelinSvanberg1975}\\
7$^2P_{3/2}$  &0.49(4)            &0.                 &" &"&~"  \\
8$^2S_{1/2}$  &5.99(8)            &-                   &"&"&~" \\
"                      &6.8(12)            &-                   &VC  &TPSDS&~\cite{ThompsonStoicheff1983} \\
"                      &6.2(12)            &-                    &VC&TPSDS&~\cite{OttoHuettner2002} \\
"                      &5.99(8)            &-                    &-&-   &Recommended \\
9$^2S_{1/2}$  & 4.0(11)           &-                    &TPSDS   &" &~\cite{OttoHuettner2002}  \\
10$^2S_{1/2}$&2.41(5)            &-                    &VC&MLC&~\cite{BelinSvanberg1975b} also\\
&            &                   &&&in~\cite{Arimondo1977}\\
 "                     &2.6(12)            &-                    &VC&TPSDS&~\cite{OttoHuettner2002} \\ 
 "                     &2.41(5)            &-                    &-&-   &Recommended \\
11$^2S_{1/2}$ & 2.1(9)            &-                    &VC&TPSDS&~\cite{OttoHuettner2002}\\
12$^2S_{1/2}$ & 1.5(12)          &-                    &"&"&~"\\ 
13$^2S_{1/2}$ & 1.9(14)         &-                      &"&"&~"\\
 14$^2S_{1/2}$ & 1.0(16)        & -                     &"&"&~"\\
\hline 
&&&{\bf $^{40}$K}&&\\ 
3$^2D_{3/2}$&|1.07(2)|      &|0.4(1)|     &VC&HQB+DD&~\cite{Sieradzan1997} \\
3$^2D_{5/2}$& |0.71(4)|      &|0.8(8)|     &"&"&"\\
4$^2S_{1/2}$&-285.7308(24)  &  -                 &AB&MWS&From~\cite{Arimondo1977} \\
4$^2P_{1/2}$&-34.49(11)        &-                   &AB&FML    &~\cite{BendaliVialle1981}\\
"                    &-34.523(25)      &-                   &AB&LIF+FC&~\cite{FalkeGrosche2006}\\
"                    &-34.523(25)      &-                   &-&-   &~Recommended \\
4$^2P_{3/2}$&-7.59(6)            &-3.5(5)          &VC&MLC&~\cite{NeySvanberg1968} \\
"                    &-7.48(6)            &-3.23(50)      &AB&FML &~\cite{BendaliVialle1981}\\
"                    &-7.585(10)        &-3.445(90)    &AB&LIF+FC&~\cite{FalkeGrosche2006}\\ 
"                    &-7.585(10)        &-3.445(90)    &-&-   &~Recommended \\
5$^2P_{1/2}$ &-12.0(9)           &-                   &VC&SAS&~\cite{BehrleKohl2011}\\ 
5$^2P_{3/2}$&-2.45(2)            &-1.16(22)      &VC&MLC&From~\cite{Arimondo1977} \\  
 \hline
&&&{\bf $^{41}$K}&&\\ 
3$^2D_{3/2}$ & |0.55(3)|        &|0.51(8)|   &VC&HQB+DD&~\cite{Sieradzan1997}\\
3$^2D_{5/2}$& |0.40(2)|        &<|0.2|        &"&"&~"\\ 
4$^2S_{1/2}$&127.0069352(6)  &   -               &AB&MWS&From~\cite{Arimondo1977}\\
"                    &126.9(8)              &-                  &AB&LIF&~\cite{TouchardVialle1982K} \\
"                    &127.0069352(6)   &-                 &-&-   &~Recommended \\
4$^2P_{1/2}$&15.19(21)            &-                  &VC&FML&~\cite{BendaliVialle1981}\\
"                    &15.1(8)                &-                  &AB&LIF&~\cite{TouchardVialle1982K}\\
"                    &15.245(42)          &-                  &AB&LIF+FC&~\cite{FalkeGrosche2006}\\
"                    &15.245(42)          &-                  &-&-   &~Recommended \\
4$^2P_{3/2}$&3.40(8)                &3.34(24)     &VC&MFD&~\cite{Ney1969}\\
"                    &3.43(5)                &0.                 &VC&MFD&~\cite{KrainskaMiszczak1981}\\
"                    &3.325(15)           &3.230(23)    &VC&HQB&~\cite{Sieradzan1995}\\
"                    &3.363(25)           &3.351(71)    &AB&LIF+FC&~\cite{FalkeGrosche2006}\\
"                    &3.342(12)           &3.242(22)    &-&-   &~w.a. \\
5$^2S_{1/2}$&30.75(75)           &   -               &VC&ORFDR&~\cite{GuptaSvanberg1973}\\
5$^2P_{1/2}$ &4.96(17)            &  -                &VC&MLC&~\cite{Ney1969} \\
5$^2P_{3/2}$&1.08(2)               &1.06(4)        &VC&MLC&~\cite{Ney1969} \\
6$^2S_{1/2}$&12.03(40)           &-                  &VC&ORFDR&~\cite{GuptaSvanberg1973}  \\
"                    &11.8(13)              &-                  &VC&TPSDS&~\cite{KiranKumar2011}, also \\ 
                    &              &                  &&&in~\cite{KiranKumar2014} \\ 
"                    &12.03(40)           &-                   &-&-   &~Recommended \\
6$^2D_{3/2}$&|0.14(2)|             &|0.05(2)|  &VC&HQB&~\cite{GlodzKrainska1985b}\\
7$^2S_{1/2}$&9.0(9)                &-                   &TD&TPSDS&~\cite{ThompsonStoicheff1983}\\ 
"                    &6.5(10)              &-                   &VC&TPSDS&~\cite{OttoHuettner2002} \\
"                    &7.9(12)                &-                   &-&-&~w.a.(w.e.e.) \\
8$^2S_{1/2}$&2.9(8)                &-                    &TD&TPSDS&~\cite{ThompsonStoicheff1983} \\
"                    &3.5(13)              &-                    &VC&TPSDS&~\cite{OttoHuettner2002} \\
"                    &3.1(7)                &-                    &-&-   &~w.a. \\
9$^2S_{1/2}$&2.6(13)              &-                    &VC&TPSDS&~\cite{OttoHuettner2002}\\
10$^2S_{1/2}$&2.0(24)            &-                    &"&"&~" \\
11$^2S_{1/2}$&2.0(11)            &-                     &"&"&~" 
\end{longtable*}

\subsection{Rubidium} 
The long list of recent spectroscopic data  for this atom is the result of the laser cooling research in worldwide spread laboratories.  For both the 85 and 87 isotopes, the ultracold atomic samples have produced very precise measurements of several hyperfine splittings, in particular for the ground state and the Rydberg ones from $n$=26 to $n$=46. Note that $^{87}$Rb has 50 neutrons, the exact magic number that makes it a closed shell.\\
\indent Entries with very high precision  and good agreement are presented in the Tables~\ref{Table:85Rb} and \ref{Table:87Rb} for $^{85}$Rb and $^{87}$Rb, respectively.  The Tables report the weighted average of the two published measured values for the $^{87}$Rb 6$^2P_{1/2}$ state  by~\cite{NyakangoPandey2020}, while for the $^{85,87}$Rb 5$^2S_{1/2}$,  and  6$^2P_{1/2}$ states, the reported average was communicated by~\cite{ShinerBernard2007}. The error bars of the $6^2S_{1/2}$ hyperfine constants for both isotopes measured by~\cite{OrsonKnize2021} and the measurements with a 30 KHz precision of~\cite{McLaughlinKnize2022}  were communicated privately by M. Lindsay. For the $A$ constants of the 57 and 58 $^2S_{1/2}$ states measured by~\cite{Meschede1987} for the $^{87}$Rb the missing error bar is assumed equal to the $^{85}$Rb ones.\\
\indent  For $n^2D$ ($n\ge6$)  states the investigations by~\cite{SvanbergTsekeris1975}, ~\cite{WijngaardenKoh1993} and by ~\cite{GlodzKrainska1987,GlodzKrainska1990,GlodzKrainska1991,KrainskaMiszczak1994} produced only the relative signs of the $A$ and $B$ constants. Their signs are determined here on the basis of the scaling laws presented in Fig. \ref{ScalingBPDstates} of  Sec.~\ref{Scalinglaws}. \\ 
\indent For the 85 isotope in Table \ref{Table:85Rb}, the  ground state hyperfine  measurement in a maser by~\cite{Tetu1976} agrees with and supersedes the atomic beam MWS results reported in~\cite{Arimondo1977}. The following Table entries  based on saturated absorption spectroscopy by~\cite{BarwoodRowley1991,ShinerBernard2007} or atomic beam laser spectroscopy by~\cite{Duong1993} have a lower precision. They  are superseded by the clock measurements in an optical fountain by~\cite{WangWang2019}. The $^{87}$Rb fountain progress was summarized in 2012 by the International Committee for Weights and Measures with their recommended Table~\ref{Table:87Rb} value. Later observations by~\cite{GuenaBize2014,Ovchinnikov2015} improved its precision.\\
 \indent Data for the $5^2P_{1/2}$ level of both isotopes are classified in two groups. The first one includes~\cite{BarwoodRowley1991}, and ~\cite{Maric2008}  while the second one includes the optical spectroscopy determination by~\cite{BeachamAndrew1971} recommended in~\cite{Arimondo1977}, the ~\cite{BanerjeeNatarajan2004} and ~\cite{DasNatarajanRb2006} data from the Bangalore research team, and the ~\cite{Rupasinghe2022} value. The agreement of the results within each group is good. However, the first group compared to the second one derives the $^{85}$Rb  $A$ constant  lower by 0.146(13) MHz, and the $^{87}$Rb $A$ value higher by 2 202(33) MHz. The data of both groups  lead to peculiar values for the hyperfine anomaly. The  higher measured values for $^{87}$Rb are closer to the theoretical predictions by~\cite{SafronovaSafronova2011} of 408.53 MHz and by~\cite{GrunefeldGinges2019} of 410.06 MHz. The  only theoretical prediction for $^{85}$Rb  by~\cite{PalPorsey2007} of 119.192 MHz is off by a few MHz above both group results.  The search for similar systematic errors as discussed  for the case of the $^{39}$K 4$^2P_{1/2}$ state combined with the present restricted data set does not resolve the discrepancy. For this state the weighted average entry in both Tables~\ref{Table:85Rb} and \ref{Table:87Rb} is reported with a large error bar determined from the $\chi^2_{\rm{red}}$ approach.\\
\indent For the $^{85}$Rb 5$^2P_{3/2}$ state, excellent agreement exists for the $A$ value. This is not the case for the  $B$ constant, where  the SAS measurement by~\cite{BarwoodRowley1991} considered as a reference point increases greatly the $\chi^2$ and as a consequence the error bar. An agreement within less than 30 kHz is reached for most data of the $^{87}$Rb 5$^2P_{3/2}$  state. An excellent  relative precision of $2\times10^{-5}$ was reached by two separate measurements on the  $^{87}$Rb isotope by~\cite{YeHall1996,DasNatarajan2008}. In Table \ref{Table:C-Constant}, the ~\cite{YeHall1996} data reexamined by ~\cite{Gerginov2009} bring evidence of the octupole contribution to the Rb hyperfine interactions.\\
 \indent  Another example of large discrepancies is found for the $^{85}$Rb  $6^2P_{1/2}$ state. The 39.470(32) MHz SAS+FC measurement by~\cite{GlaserFortagh2020} presents a difference exceeding the error bar, compared to the values reported by by~\cite{ShinerBernard2007}  based on the same SAS+FC technique, and also the ORFDR measurement by~\cite{FeiertagzuPutlitz1973} and the OODR-EIT  measurement by~\cite{NyakangoPandey2020}. Our derivation of the hyperfine splitting from the measured optical frequencies of~\cite{GlaserFortagh2020} leads to the  39.11(22) MHz value in good agreement with other ones.  Therefore, the ~\cite{GlaserFortagh2020} entry is not included in the weighted average. \\
  \indent  In contrast, the $6^2P_{3/2}$ Glaser 
 et al. (2020) data for both isotopes are in very good agreement with the earlier radiofrequency and level-crossing data. For that state in $^{85}$Rb, the $A$ value derived by~\cite{ZhangRb2017}  on the basis of saturated absorption and EIT measurements is close to those of other references. However, that experiment produced a $B$ constant with a large deviation from other values, probably because the hyperfine lines were not well resolved. Both their $A$ and $B$ values are not included into the w.a.. \\
 \indent The $7^2S_{1/2}$ state of both isotopes has received a wide attention because of the large probability for the two-photon excitation from the ground state. Precision at the $1\times10^{-6}$ level is reached in the 87 isotope and at the $2\times10^{-5}$ level in the 85 one, limited by the 2 kHz resolution of the frequency comb in~\cite{KrishnaNatarajan2005,Chui2005,BarmesEikema2013,Morzynski2013}. However, in both isotopes the EIT results by~\cite{KrishnaNatarajan2005} are lower than the other ones by $\approx 400$ kHz, while  their claimed precision is $\approx 20$ kHz precision. The presence of light shifts originated by the intense control laser producing the EIT signal was not tested. The weighted average are performed excluding the ~\cite{KrishnaNatarajan2005} values.\\
 \indent For the $n=(9-13)$  $^2S_{1/2}$ states, the optical spectra recorded by~\cite{StoicheffWeinberger1979} produce hyperfine constants in good agreement with more recent ones, but only for the 85 isotope. Their $n=(9-11)$ data for the 87 isotope are very far off and not included in the w.a..\\
\begin{longtable*}{cddccl}
\caption{Measured $A$ and $B$ values for $^{85}$Rb isotope\label{Table:85Rb}}\\
\hline
State& A (MHz) & B(MHz)&Sample&Technique  &Ref.   \\
\hline
\endfirsthead
\multicolumn{6}{c}%
{\tablename\ \thetable\ -- \textit{Continued from previous page}} \\
\hline
State& A (MHz) & B(MHz)&Sample&Technique  &Ref.   \\
\hline
\endhead
\hline \multicolumn{6}{r}{\textit{Continued on next page}} \\
\endfoot
\hline
\endlastfoot
4$^2D_{3/2}$&7.3(5)                       &0.               &VC&ORFDR&~\cite{LamHapper1980} \\
"                    &7.329(35)                 &4.52(23)     & VC&OODR&~\cite{MoonLeeSuh2009} \\
"                    &7.329(35))                &4.52(23)     &-&-&~Recommended\\
4$^2D_{5/2}$&-5.2(3)                      &0.               &VC&ORFDR&~\cite{LamHapper1980} \\
"                    &-5.06(10)                  &7.42(15)     &MOT&OODR &~\cite{SinclairDuxbury1994} \\
"                    &-4.978(4)                  &6.560(52)   &VC&OODR+EIT&~\cite{WangWang2014Rb} \\
"                    &-5.008(9)                  &7.15(15)     &VC&OODR+FC  &~\cite{LeeMoon2015} \\
"                    &-4.983(7)                  &6.70(21)    &-&-& ~w.a.(w.e.e.)\\
5$^2S_{1/2}$&1011.910813(2)       &-                  &AB&MA&~\cite{Tetu1976} \\
"                    &1011.894(9)              &-                 &VC&SAS&~\cite{BarwoodRowley1991}\\
"                    &1011.9108(3)            &-                  &AB&LIF&~\cite{Duong1993} \\
"                    &1011.914(12)            &-                  &VC&SAS+FC&~\cite{ShinerBernard2007}\\
"                    &1011.9108149406(1)&-                  &FOUNT&MWS&~\cite{WangWang2019}\\
"                    &1011.9108149406(1)&-                   &-&-&~Recommended\\
5$^2P_{1/2}$&120.72(25)               &-                    &VC&OS&~\cite{BeachamAndrew1971}\\
"                    &120.499(10)             &-                   &VC&SAS&~\cite{BarwoodRowley1991}\\
"                    &120.64(2)                  &-                    &VC&SAS&~\cite{BanerjeeNatarajan2004}\\
"                    &120.645(5)                &-                    &VC&SAS&~\cite{DasNatarajanRb2006} \\ 
"                    &120.500(13)              &-                   &MOT&LIF+FC&~\cite{Maric2008} \\
"                    &120.79(29)                &-                   &VC&SAS&~\cite{Rupasinghe2022} \\
"                    &120.605(29)                &-                   &-&-&~w.a.(w.e.e.)\\
5$^2P_{3/2}$&25.009(22)                &25.88(3)       &VC&ORFDR&From~\cite{Arimondo1977}\\
"                    &25.3 (4)                     &21.4 (40)      &AB&LIF&~\cite{ThibaultHuber1981} \\
"                    &24.988 (31)               &25.693 (31)  &VC&SAS&~\cite{BarwoodRowley1991} \\
"                    &25.038(5)                  &26.011(22)   &VC&SAS&~\cite{Rapol2003} \\
"                    &25.041(6)                  &26.013(25)   &VC&SAS&~\cite{BanerjeeNatarajan2003} \\
"                    &25.0403(11)              &26.0084(49) &VC&SAS&~\cite{DasNatarajan2008} \\
"                    &25.0401(11)                  &26.000(22) &-&-&~ w.a.; w.a.(w.e.e.).\\
5$^2D_{3/2}$&4.2699(2)                 &1.9106(8)    &VC&TPSDS&~\cite{NezMillerioux1994}\\
"                    &4.43(28)                    &1.7(24)         &MOT&OODR+RIS&~\cite{GabbaniniLucchesini1999} \\
"                    &4.2699(2)                  &1.9106(8)     &-&-&~Recommended \\
5$^2D_{5/2}$&-2.2112(12)              &2.6804(200) &VC&TPSDS&~\cite{NezMillerioux1994}\\
"                     &-2.196(52)               &2.51(53)       &VC&TPSDS&~\cite{Grove1995} \\
"                     &-2.31(23)                 &2.7(27)         &MOT&OODR+RIS&~\cite{GabbaniniLucchesini1999} \\
"                     &-2.222 (19)              &2.664(130)   &VC&EIT&~\cite{YangWang2017} \\
"                     &-2.2112(12)             &2.6804(200)  &-&-&~Recommended\\
6$^2S_{1/2}$ & 239.18(3)               &-                    &VC&FML& ~\cite{PerezGalvanSprouse2007} also\\
                       &                                &                     &     &.      & in~\cite{PerezGalvanOrozco2008}  \\ 
"                     &234.(30)                    &-                   &VC&TPSDS&\cite{OrsonKnize2021}\\
"                     &239.057(10)             &-                   &VC&TPSDS&\cite{McLaughlinKnize2022}\\
"                     &239.069(26)             &-                   &-&-&~w.a.(w.e.e.)\\
6$^2P_{1/2}$ &39.11(3)                   &-                   &VC&ORFDR&~\cite{FeiertagzuPutlitz1973}  \\
"                     &39.123(9)                 &-                   &VC&SAS+FC&~\cite{ShinerBernard2007}\\
"                     &39.470(32)               &-                   &VC&SAS+FC&~\cite{GlaserFortagh2020}  \\
"                     &39.122(9)                 &-                   &-&-&~See text. w.a. \\
6$^2P_{3/2}$ &8.179(12)                 &8.190(49)  &VC&ORFDR+MLC &From~\cite{Arimondo1977} \\
"                     &8.220(3)                   &5.148(3)    &VC&SAS&~\cite{ZhangRb2017}  \\
"                     &8.1667(94)               &8.126(54)  &VC&SAS+FC&~\cite{GlaserFortagh2020}  \\
"                     &8.171(7)                   &8.161(36)   &-&-& ~See text. w.a. \\
6$^2D_{3/2}$ &2.28(6)                     &0.                &VC&MLC&~\cite{HogervostSvanberg1975} \\
"                     &2.32(6)                     &1.62(6)        &VC&HQB&~\cite{WijngaardenHapper1986} \\
"                     &2.30(4)                     &1.62(6)    &-&-&~w.a.\\
6$^2D_{5/2}$ &-1.069(18)                &-0.41(41)    &VC&OODR&~\cite{BrandenbergerLindley2015} \\
7$^2S_{1/2}$ &94.7(1)                     &-             &MOT&TPSDS &~\cite{SnaddenFerguson1996} \\
"                     &94.2(6)                      &-             &MOT&OODR   &~\cite{Gomez2004} \\
"                     &94.085(18)                &-             &VC&EIT  &~\cite{KrishnaNatarajan2005} \\
"                     &94.658(19)                &-             &VC&TPSDS+FC &~\cite{Chui2005} \\
"                     &94.6807(37)              &-             &"&" &~\cite{BarmesEikema2013}\\
"                     &94.6784(23)              &-             &"&" &~\cite{Morzynski2013,Morzyski2014}\\
"                     &94.684(2)                  &-             &"&"&~\cite{MorgenwegEikema2014} \\
"                     &94.6813(15)                    &-              &-&-&~See text. w.a. \\
7$^2P_{1/2}$ &17.68(8)                    &-             &VC&ORFDR&~\cite{FeiertagzuPutlitz1973}  \\
7$^2P_{3/2}$ &3.71(1)                      &3.68(8)   &"&"&~\cite{BuckaMinor1961} \\
7$^2D_{3/2}$ &1.34(1)                      &0.            &VC&MLC &~\cite{HogervostSvanberg1975} \\
"                      &1.415(30)                  &0.31(6)    &VC&HQB &~\cite{WijngaardenSagle1991}\\
"                      &1.40(10)                    &0.             &VC&TPSDS&~\cite{OttoHuettner2002} \\
"                      &1.35(2)                      &0.31(6)    &-&-&~w.a.(w.e.e.)\\
7$^2D_{5/2}$ &-0.55(10)                   &0.             &VC&MLC&~\cite{HogervostSvanberg1975} \\
8$^2S_{1/2}$ &45.2(20)                     &-               & VC&ORFDR&~\cite{GuptaSvanberg1973} \\
"                     &47.1(20)                     &-               &VC&TPSDS&~\cite{OttoHuettner2002} \\
"                     &46.1(14)                     &-               &-&-&~w.a. \\
8$^2P_{3/2}$ &1.99(2)                       &1.98(12)   & VC&ORFDR&~\cite{zuPutlitz1968} \\
8$^2D_{3/2}$ &0.84(1)                       &0.              &VC&MLC &~\cite{HogervostSvanberg1975}\\
"                      &0.879(8)                     &0.15(2)     &VC&HQB &~\cite{WijngaardenKoh1993}\\
"                      &0.864(19)                     &0.15(2)     &-&-&~w.a. (w.e.e.); w.a. \\
8$^2D_{5/2}$ &-0.35(7)                       &0.             &VC&MLC&~\cite{HogervostSvanberg1975}\\
9$^2S_{1/2}$ &30.(2)                           & -              &TD&TPSDS&~\cite{StoicheffWeinberger1979} \\
"                      &33.5(15)                      & -              &VC&TPSDS&~\cite{OttoHuettner2002}\\
"                      &32.2(12)                     &-                &-&-&~w.a. \\
9$^2D_{3/2}$ &0.561(11)                    &0.20(3) &VC &HQB&~\cite{KrainskaMiszczak1994} \\ 
10$^2S_{1/2}$&23.(3)                          & -              &TD&TPSDS&~\cite{StoicheffWeinberger1979} \\
"                      &22.2(16)                       & -               &VC&TPSDS&~\cite{OttoHuettner2002} \\
"                      &22.4(14)                       &-    &-&-&~w.a.  \\
10$^2D_{3/2}$&0.393(8)                 &0.141(13) &VC&HQB&~\cite{GlodzKrainska1993}\\
11$^2S_{1/2}$&13.(6)                      & -              &TD&TPSDS&~\cite{StoicheffWeinberger1979} \\
"                      &17.1(19)                     &-                &VC&TPSDS&~\cite{OttoHuettner2002} \\
"                      &16.7(18)                      &-    &-&-&~w.a.  \\
11$^2D_{3/2}$ &   0.283(6)                 &0.100(11) &VC&HQB&~\cite{GlodzKrainska1993} \\
12$^2S_{1/2}$ &7.3(2)                      & -              &TD&TPSDS&~\cite{StoicheffWeinberger1979} \\
13$^2S_{1/2}$ &8.3(4)                     & -              &"&"&~" \\
28$^2S_{1/2}$ &0.322(26)               &-          &MOT&MWS&~\cite{LiGallagher2003} \\
29$^2S_{1/2}$ &0.280(26)               &-          &"&" &~" \\
30$^2S_{1/2}$ &0.224(20)               &-          &"&"&~" \\
31$^2S_{1/2}$ &0.252(31)                &-          &"&" &~" \\
32$^2S_{1/2}$ &0.209(22)                &-          &"&" &~"\\
33$^2S_{1/2}$ &0.182(21)                &-          &"&" &~"\\
43$^2S_{1/2}$ &0.0804(2)                &-          &MOT&MWS &~\cite{RamosRaithel2019}\\
44$^2S_{1/2}$ &0.0743(3)                &-          &"&" &~"\\
45$^2S_{1/2}$ &0.0703(7)                &-          &"&" &~"\\
46$^2S_{1/2}$ &0.0653(1)                &-          &"&" &~"\\
50$^2S_{1/2}$ &0.057(13)                &-          &AB&MWS &\cite{Meschede1987}\\
51$^2S_{1/2}$ &0.057(12)                &-          &"&" &~"\\
52$^2S_{1/2}$ &0.053(10)                &-          &"&" &~"\\
53$^2S_{1/2}$ &0.050(10)                &-          &"&" &~"\\
54$^2S_{1/2}$ &0.050(10)                &-          &"&" &~"\\
55$^2S_{1/2}$ &0.047(10)                &-          &"&" &~"\\
56$^2S_{1/2}$ &0.041(8)                  &-          &"&" &~"\\
57$^2S_{1/2}$ &0.040(7)                  &-          &"&" &~"\\
58$^2S_{1/2}$ &0.036(7)                  &-          &"&" &~"\\
59$^2S_{1/2}$ &0.036(3)                  &-          &"&" &~"\\
\end{longtable*}
\begin{longtable*}{cddccl}
\caption{Measured $A$ and $B$ values for $^{87}$Rb isotope\label{Table:87Rb}}\\
\hline
State& A (MHz) & B(MHz)&Sample&Technique  &Ref.  \\
\hline
\endfirsthead
\multicolumn{6}{c}%
{\tablename\ \thetable\ -- \textit{Continued from previous page}} \\
\hline
State& A (MHz) & B(MHz)&Sample&Technique  &Ref.  \\
\hline
\endhead
\hline \multicolumn{6}{r}{\textit{Continued on next page}} \\
\endfoot
\hline
\endlastfoot
4$^2D_{3/2}$ &25.1(9)                               &0.                &VC&ORFDR&~\cite{LamHapper1980} \\
"                  &24.75(12)                           &2.19(11)      &VC&OODR&~\cite{MoonLeeSuh2009} \\
"                  &24.75(12)                           &2.19(11)      &-&-&~Recommended\\
4$^2D_{5/2}$  &-16.9(6)                         &0.                &VC&ORFDR&~\cite{LamHapper1980} \\
"                   &-16.747(10)                      &4.149(59)  &VC&SAS+FC &~\cite{LeeMoon2007,LeeMoon2007er} \\
"                   &-16.801(5)                        &3.645(30)   &    VC&OODR+EIT &~\cite{WangWang2014Rb}  \\
"                   &-16.779(6)                        &4.112(52)    &    VC&OODR+FC  &~\cite{LeeMoon2015} \\
"                   &-16.786(10)                      &3.82(16)     &-&-&~w.a.(w.e.e.)\\
5$^2S_{1/2}$&3417.330 (7)                  &-                   &VC&SAS&~\cite{BarwoodRowley1991} \\
"                    &3417.3415(5)                 &-                    &AB&LIF&~\cite{Duong1993} \\ 
"                    &3417.341305452156(4)  &-                   &-&-&~\cite{CCTF2012}\\
"                    &3417.353(19)                  &-                   &VC&SAS+FC&~\cite{ShinerBernard2007}\\
"                    &3417.341305452156(3)  &-                    &MOT&FOUNT&~\cite{GuenaBize2014}\\          
"                    &3417.341305452154(2)  &-                    &MOT&FOUNT&~\cite{Ovchinnikov2015} \\
"                    &3417.3413054521548(15)&-                  &-&-&~w.a.\\
5$^2P_{1/2}$ &406.2(8)                         &-                     &VC&OS&~\cite{BeachamAndrew1971}\\
"                     &408.328(15)                    &-                    &VC&SAS&~\cite{BarwoodRowley1991}\\
"                     &406.147(15)                   &-                     &VC&SAS&~\cite{BanerjeeNatarajan2004}\\
"                     &406.119(7)                     &-                     &VC&SAS&~\cite{DasNatarajanRb2006}\\
"                     &408.330(56)                   &-                     &MOT&LIF+FC&~\cite{Maric2008} \\
"                     &408.3(1)                         &-                     &ODT&OS&~\cite{NeuznerRitter2015} \\
"                    &407.75(50)                      &-                   &VC&SAS&~\cite{Rupasinghe2022} \\
"                    &406.48(33)                     &-                     &-&-&~w.a.(w.e.e.)\\
5$^2P_{3/2}$&84.29(50)                        &12.2(20)         &VC&SAS&~\cite{ThibaultHuber1981} \\
"                    &84.676 (28)                     &12.475(28)     &VC&SAS&~\cite{BarwoodRowley1991} \\
"                    &84.7185(20)                    &12.4965(37)   &MOT&SAS&~\cite{YeHall1996} \\
"                    &84.7189(22)                    &12.4942(43)   &MOT&SAS&~\cite{Gerginov2009} \\
"                    &84.7200(16)                    &12.4970(35)    &VC&SAS&~\cite{DasNatarajan2008} \\
"                    &84.745(6)                        &12.528(10)     &VC&SAS&~\cite{ChangMobarhan2017} \\
"                    &84.720(3)                        &12.497(2)       &-&-&~w.a.(w.e.e.)\\
5$^2D_{3/2}$&14.4303(5)                     &0.9320(17)   &VC&TPSDS&~\cite{NezMillerioux1994}\\
"                    &14.64(30)                        &0.8(8)       &MOT&OODR+RIS&~\cite{GabbaniniLucchesini1999}\\
"                    &14.4303(5)                      &0.9320(17)     &-&-&~Recommended\\
5$^2D_{5/2}$&-7.4605(3)                       &1.2713(20)   &VC&TPSDS&~\cite{NezMillerioux1994}\\
"                    &-7.45(21)                         &0.462(1088)  &MOT&OODR&~\cite{Grove1995}\\
"                    &-7.51(28)                         &2.7(24)    &MOT&OODR+RIS&~\cite{GabbaniniLucchesini1999} \\
"                   &-7.4605(3)                        &1.2713(20)    &-&-&~Recommended\\
6$^2S_{1/2}$&807.66(8)                        &-                    &VC&FML& 
~\cite{PerezGalvanSprouse2007} also \\ 
                      &                                       &                      &     & &in~\cite{PerezGalvanOrozco2008} \\ 
                     &797.(30)                       &-                   &VC&TPSDS&\cite{OrsonKnize2021}\\
"                    &807.341(15)                 &-                   &"&"&\cite{McLaughlinKnize2022}\\
"                    &807.35(4)                     &-          &-&-&~w.a.(w.e.e.)\\
6$^2P_{1/2}$ &132.56(3)                    &-     &VC&ORFDR&~\cite{FeiertagzuPutlitz1973}\\
"                     &132.559(13)                &-      &VC&SAS+FC&~\cite{ShinerBernard2007}\\
"                     &132.583(141)              &-&VC&OODR+EIT&~\cite{NyakangoPandey2020} \\
"                     &133.24(28)                   &-&VC&SAS+FC&~\cite{GlaserFortagh2020}\\
"                     &132.569(9)                   &-                   &-&-&~See text. w.a. \\
6$^2P_{3/2}$&27.700(17)                    &3.953(24)     &VC&ORFDR&From~\cite{Arimondo1977} \\
"                      &27.710(15)                  &4.030(42)      &VC&SAS+FC&~\cite{GlaserFortagh2020}\\
"                      &27.706(11)                  &3.972(33)      &-&-&~w.a.(w.e.e. for $B$) \\
6$^2D_{3/2}$&7.84(5)                          &0.53(6)          &VC&MLC&~\cite{SvanbergTsekeris1975}\\ 
6$^2D_{5/2}$&- 3.4(5)                          &0.               &VC&ORFDR+MLC&~\cite{HogervostSvanberg1975} \\
 "                    &-3.61(6)                         &-0.20(20)      &VC&OODR&~\cite{BrandenbergerLindley2015}\\
"                     &-3.61(6)                         &-0.20(20)       &-&-&~Recommended \\
7$^2S_{1/2}$ &319.7(1)                        &-                    &MOT&TPSDS &~\cite{SnaddenFerguson1996}\\
"                     &319.174(45)                  &-                     &VC& EIT &~\cite{KrishnaNatarajan2005} \\
"                     &319.702(65)                  &-                     &MOT&TPSDS+FC &~\cite{MarianYe2005} \\
"                     &319.759(28)                  &-                     &VC&" &~\cite{Chui2005} \\
"                     &319.7518(51)                &-                  &VC&" &~\cite{BarmesEikema2013} \\
"                     &319.7479(23)                &-            &VC&" &~\cite{Morzynski2013} also\\
                    &                &           && &in~\cite{Morzyski2014}\\
"                     &319.762(6)                    &-                   &VC&"&~\cite{MorgenwegEikema2014} \\
"                     & 319.7500(20)                &-                  &-&-&~See text. w.a. \\
7$^2P_{1/2}$ &59.92(9)                         &-                  &VC&ORFDR&~\cite{FeiertagzuPutlitz1973} \\
7$^2P_{3/2}$ &12.57(1)                        &1.762(16)  &VC&ORFDR+MLC&From~\cite{Arimondo1977} \\
7$^2D_{3/2}$ &4.53(3)                         &0.26(4)  &    VC&MLC&~\cite{SvanbergTsekeris1975}  \\
"                     &4.69(23)                        &0.          &         VC&TPSDS&~\cite{OttoHuettner2002} \\
"                     &4.53(3)                          &0.26(4)         &-&-&~Recommended \\
7$^2D_{5/2}$ &-2.0(3)                          &0.           &    VC&ORFDR+MLC&~\cite{HogervostSvanberg1975}\\
"                     &-1.85(80)                       &0.           &         VC&TPSDS&~\cite{OttoHuettner2002} \\
"                     &-1.98(28)                       &0.            &-&-&~w.a. \\
 8$^2S_{1/2}$&159.2(15)                     &-                  &VC&ORFDR&~\cite{TsekerisGupta1975} \\
 "                    &159.3(30)                      &-                  &VC&TPSDS&~\cite{OttoHuettner2002} \\
"                     &159.2(13)                      &-                  &-&-&~w.a. \\
8$^2P_{1/2}$ &32.12(11)                      &-                   &VC&ORFDR& ~\cite{TsekerisFarleyGupta1975} \\
8$^2P_{3/2}$ &6.739(15)                      &0.935(22) &VC&ORFDR+MLC& From~\cite{Arimondo1977} \\
8$^2D_{3/2}$&2.840(15)                       &0.17(2)          &VC&MLC+ORFDR&~\cite{BelinSvanberg1976a}\\
8$^2D_{5/2}$&-1.20(15)                        &0.              &VC&ORFDR+MLC& ~\cite{HogervostSvanberg1975}\\
"                     &-1.00(13)                        &0.                  &VC&TPSDS&~\cite{OttoHuettner2002} \\
"                     &-1.09(10)                        &0.                   &-&-&~w.a. \\
9$^2S_{1/2}$&90.9(8)                            &-                    &VC&ORFDR&~\cite{TsekerisGupta1975} \\
"                    &106.(3)                            &-                    &TD&TPSDS&~\cite{StoicheffWeinberger1979} \\
"                    &91.6(47)                          &-                     &VC&TPSDS&~\cite{OttoHuettner2002} \\
"                    &90.9(8)                            &-                    &-&-&~Recommended \\
9$^2P_{3/2}$&4.05(3)                             &0.55(3)         &VC&ORFDR&~\cite{BelinSvanberg1976a} \\
9$^2D_{3/2}$&1.90(1)                             &0.11(3)         &VC&MLC+ORFDR&~\cite{BelinSvanberg1976a} \\
"                    &2.01(17)                           &0.                 &VC&TPSDS&~\cite{OttoHuettner2002} \\
"                    &1.90(1)                             &0.11(3)          &-&-&~Recommended \\
9$^2D_{5/2}$&-0.80(15)                          &0.                  &VC&MLC+ORFDR&~\cite{BelinSvanberg1976a} \\
"                     &-0.740(12)                       &0.160(15)      &VC&HQB &~\cite{GlodzKrainska1990} \\  
"                     &-0.740(12)                       &0.160(15)      &-&-&Recommended \\
10$^2S_{1/2}$&56.3(2)                          &-                    &VC&ORFDR&~\cite{Farley1977} \\
"                      &70.(3)                              &-                   &TD&TPSDS&~\cite{StoicheffWeinberger1979} \\
"                      &56.1(23)                          &                     &VC&TPSDS&~\cite{OttoHuettner2002} \\
"                      &56.3(2)                            &-                   &-&-&~Recommended\\
10$^2P_{3/2}$& 2.60(8)                           &0.                  &VC&ORFDR& ~\cite{BelinSvanberg1976a}\\       
10$^2D_{3/2}$ &1.315(17)                        &0.070(11)       &VC&HQB &~\cite{GlodzKrainska1991}\\
10$^2D_{5/2}$ &-0.510(10)                       &0.098(11)        &"&" &~\cite{GlodzKrainska1987} \\
11$^2S_{1/2}$ & 37.4(3)                            &-                      &VC&ORFDR&~\cite{Farley1977}  \\
 "                      &54.(10)                         &-                       &TD&TPSDS&~\cite{StoicheffWeinberger1979} \\
 "                      &37.2(35)                       &                       &VC&TPSDS&~\cite{OttoHuettner2002} \\
"                       &37.4(3)                         &-                   &-&-&~Recommended\\
11$^2D_{3/2}$ &0.955(11)                      &0.049(6)         &VC&HQB &~\cite{GlodzKrainska1991}\\
11$^2D_{5/2}$ &-0.361(7)                       &0.071(11)         &"&" &~\cite{GlodzKrainska1989} \\ 
12$^2S_{1/2}$  &27.(8)                           &-                      &TD&TPSDS&~\cite{StoicheffWeinberger1979} \\
12$^2D_{3/2}$ &0.715(12)                      &0.037(8)            &VC&HQB &~\cite{GlodzKrainska1991}\\
12$^2D_{5/2}$ &-0.266(9)                       &0.063(14)       &"&"&~\cite{GlodzKrainska1989}\\ 
13$^2S_{1/2}$  &23.(8)                            &-                    &TD&TPSDS&~\cite{StoicheffWeinberger1979} \\ 
13$^2D_{5/2}$ &-0.20(1)                          &0.05(2)           &VC&HQB&~\cite{GlodzKrainska1989}\\
20$^2S_{1/2}$ & 3.891(2)                         &-                     &VC&EIT &~\cite{TauschinskySpreeuw2013} \\
21$^2S_{1/2}$ &3.249(2)                          &-                     &"&" &~" \\
22$^2S_{1/2}$ &2.721(3)                          &-                     &"&" &~" \\
23$^2S_{1/2}$ & 2.390(4)                          &-                     &"&" &~" \\
24$^2S_{1/2}$ & 2.115(5)                           &-                     &"&" &~" \\
28$^2S_{1/2}$ &1.07(5)                              &-                     &MOT&MWS &~\cite{LiGallagher2003} \\
29$^2S_{1/2}$ &0.97(5)                              &-                   &"&" &~" \\
30$^2S_{1/2}$ &0.78(4)                              &-                    &"&" &~" \\
31$^2S_{1/2}$ &0.81(7)                              &-                   &"&"&~" \\
32$^2S_{1/2}$ &0.71(5)                              &-                    &"&" &~" \\
33$^2S_{1/2}$ &0.63(4)                              &-                    &"&" &~" \\
50$^2S_{1/2}$ &0.185(20)                          &-                    &AB&MWS&\cite{Meschede1987}\\
51$^2S_{1/2}$ &0.170(18)                   &-          &"&" &~"\\
52$^2S_{1/2}$ &0.165(18)                   &-          &"&" &~"\\
53$^2S_{1/2}$ &0.160(10)                   &-          &"&" &~"\\
54$^2S_{1/2}$ &0.145(18)                   &-          &"&" &~"\\
55$^2S_{1/2}$ &0.145(15)                   &-          &"&" &~"\\
56$^2S_{1/2}$ &0.142(13)                   &-          &"&" &~"\\
57$^2S_{1/2}$ &0.135(13)                   &-          &"&" &~"\\
58$^2S_{1/2}$ &0.111(13)                   &-          &"&" &~"\\
59$^2S_{1/2}$ &0.105(10)                      &-         &"&" &~"\\
\end{longtable*}
\begin{table*}
\caption{Measured $C$ values for $^{87}$Rb and $^{133}$Cs}
\label{Table:C-Constant}
\begin{tabular}{ccccl}
\hline
\hline
State& C(kHz)&Sample&Technique  &Ref.  \\
\hline
&&$^{87}$Rb&\\ 
5$^2P_{3/2}$&-0.12(9)   &MOT&SAS&~\cite{Gerginov2009} \\
&&$^{133}$Cs&\\
6$^2P_{3/2}$&0.56(7) &AB&LIF&~\cite{GerginovTanner2003} \\
"                  &0.87(32)& VC &CCS& ~\cite{DasNatarajan2005} \\
6$^2D_{3/2}$&  4.3(10)       &VC &TPSDS&~\cite{ChenCheng2018} \\
\hline
\hline
\end{tabular}
\end{table*}

\subsection{Cesium} 
Considering the large set of measured  values that cover up Rydberg states with high $n$ numbers,  cesium is a favourite atom for hyperfine spectroscopy.  In addition. there is a good   (or very good) agreement for the large majority of states. \\
\indent The use of frequency combs in order to perform  absolute optical frequency measurements produces very precise values for the explored states. For instance, the $6^2P_{1/2}$ $A$  constant was reported with a $\approx3\times10^{-4}$ precision  in~\cite{Arimondo1977}, while it reached $\approx3\times10^{-6}$ in~\cite{Gerginov2006}. A similar spectacular improvement is associated with the 8$^2S_{1/2}$ state, object of several investigations owing to its large two-photon excitation probability.\\
\indent The nuclear magnetic octupole dipole moment was measured for the first time  in the $6^2P_{3/2}$ state by~\cite{GerginovTanner2003} with the value of $C=0.56(7)$ kHz, and remeasured as $C=0.87(32)$ kHz by~\cite{DasNatarajan2008}, as shown in Table~\ref{Table:C-Constant}. The first reference reaches a higher precision, but  leads to a $B$ constant in poor agreement with the values by the second reference and by ~\cite{TannerWieman1988}. For the 6$^2D_{3/2}$ state, the octupole moment was recently measured by~\cite{ChenCheng2018}  with the value of $C=4.3(10)$ kHz, much larger than the above value for the $6^2P_{3/2}$ state. \\
\indent  Off-diagonal elements between 6$P_{1/2}$ and 6$P_{3/2}$ levels derived theoretically in~\cite{JohnsonDerevianko2004} at the level of $\approx 40$ Hz,  are negligible even at the high precision level of the 6$P_{3/2}$  state hyperfine data.\\
\indent For several high $nD$, with $n\ge10$, states, the investigations by ~\cite{SvanbergBelin1974,BelinSvanberg1976b}, ~\cite{DeechSeries1977,NakayamaSeries1981}, ~\cite{SagleWijngaarden1991}, ~\cite{GlodzKrainska1987,GlodzKrainska1990,GlodzKrainska1991} produced only the absolute value of the $A$ constant. Their signs are determined here on the basis of the scaling laws. All $B$ values for those states were assumed equal zero.\\
\indent The  inversion of the $^2D_{5/2}$  hyperfine states is basically due to core-polarization and electron-correlation effects induced by the valence electron, as initially pointed out by~\cite{FredrikssonSvanberg1980}  and carefully examined recently in~\cite{Auzinsh2007,GrunefeldGinges2019,TangShi2019}. \\
\indent The  6$^2S_{1/2}$  $A$ coefficient corresponding to the ground state hyperfine splitting is not listed in the Table~\ref{Table:Cs} because it is related, as the 2298.157943 MHz frequency, to the Bureau International des Poids et Mesures definition of the second.\\
\indent For the 6$^2D_{3/2}$ state, the Table~\ref{Table:Cs} reports the  value by~\cite{Kortyna2006}, because the  remeasured value in~\cite{KortynaSafronova2011} using a different method is less precise. The $A$ and $B$ values measured by~\cite{ChengHsu2017} are not included into the calculation of the weighted average because their fit does not reproduce all the measured hyperfine frequencies within the reported error bar.  \\
\indent For the 7$^2D_{5/2}$ state, several $A$ values were measured with good overall agreement. This is not the case for the $B$ values, where large discrepancies are reported. The small values of the hyperfine constants limit the frequency resolution. Three experiments~\cite{LeeChui2011,WangWang2020,WangJia2021} measured a restricted number of hyperfine splittings with a limited agreement of their frequency. ~\cite{StalnakerTanner2010} reported a full high-resolution spectrum and  a careful study of systematic errors. Having observed 
a dependence of  the $B$ value on the applied magnetic field,  they increase the error bar of their $B$ measurement in order to cover both negative and positive values.   Their $A$ and $B$ values are recommended in Table~\ref{Table:Cs}.\\
\begin{figure}
\centering
\includegraphics[width= 1.0\columnwidth]{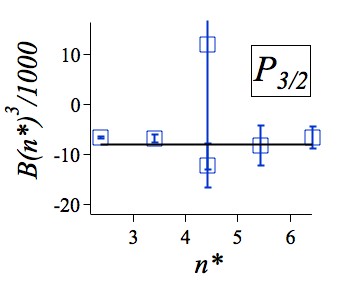}
\caption{ For Cs $^2P_{3/2}$$B(n^*)^3$ scaling test, with $B$ in MHz,  vs. the $n$ number in logarithmic scale  states. For the $n=8$ state one negative and one positive value are reported here from Table~\ref{Table:Cs}.  The quantum defect parameters are derived from~\cite{LorenzenNiemax1984}.  The continuous horizontal line reporesents a fit of the negative values based on the $1/(n^*)^3$  scaling. }
\label{BScaling_P32}
\end{figure}
\indent For the 8$^2P_{3/2}$ state, the $A$ and $B$  values  examined by~\cite{Arimondo1977}   were based on early ORFDR investigations by \cite{BarbeyGeneux1962,BuckavonOppen1963,FaistKoide1964}, all suffering from radiofrequency shifts as discussed in this last reference. All of them are reported in Table~\ref{Table:Cs}. \cite{Arimondo1977} recommended the~\cite{FaistKoide1964} values, where the shift corrections were included. That $A$ value  and the one by ~\cite{Bayram2014} agree at 0.15 MHz level. Table~\ref{Table:Cs} reports the w.a. of  ~\cite{FaistKoide1964} and ~\cite{Bayram2014}  with $\chi^2_{red}$ correction. The $B$ values by ~\cite{FaistKoide1964} and ~\cite{Bayram2014} are identical except for their sign. ~\cite{Bayram2014}  defended their positive value on the basis of their earlier Na 3$P_{3/2}$ hyperfine constants by~\cite{YeiSieradzan1993} using a similar technique and  agreeing with the best other measurements as in Table~\ref{Table:Na}.  A negative $B$ value is confirmed by the scaling law of Fig.~\ref{BScaling_P32}. Despite weighted  average being -0.12(5), Table~\ref{Table:Cs} recommends the value of~\cite{FaistKoide1964}.\\
\indent For the measured constant of 9$^2S_{1/2}$,  the high reported precision of~\cite{JinJia2013} leads to an anomalously large contribution to $\chi^2_{red}$.  When that value is excluded from the w.a., the error bar is greatly reduced. It is worth noting that  the 8$^2S_{1/2}$ and 7$^2D_{3/2}$ values presented by the same authors with a similar precision match very well those by other authors. The 9$^2S_{1/2}$ recommended value is the    
~\cite{MorgenwegEikema2014} measurement having a $2\cdot 10^{-5}$ precision. \\
   \indent The 9 and 10 $^2P_{3/2}$ values  by ~\cite{RydbergSvanberg1972} are corrected for the $g_j$ value in ~\cite{Arimondo1977}.   ~\cite{RydbergSvanberg1972} derive the 10$^2P_{3/2}$ $B$ value on the basis of the average measured ratio $B/A=-0.010(3)$ in the lower $n$ $^2P_{3/2}$ levels.\\
\indent  We have received privately from J. Deiglmayr the $A$ values for the $^2S_{1/2},^2P_{1/2}$ states with $n$ between 43 and 81 measured by~\cite{Sassmannshausen2013}.  \\ 
\begin{longtable*}{cddccl}
\caption{Measured $A$ and $B$  values for $^{133}$Cs\label{Table:Cs}}\\
\hline
State& A (MHz) & B(MHz)&Sample&Technique  &Ref.   \\
\hline
\endfirsthead
\multicolumn{6}{c}%
{\tablename\ \thetable\ -- \textit{Continued from previous page}} \\
\hline
State& A (MHz) & B(MHz)&Sample&Technique  &Ref.   \\
\hline
\endhead
\hline \multicolumn{6}{r}{\textit{Continued on next page}} \\
\endfoot
\hline
\endlastfoot
5$^2D_{3/2}$&48.6(2)         & 0.0(8)       &AB&LIF&~\cite{FredrikssonSvanberg1980}\\
 "                    &48.8(3)         & 0.             &VC&HQB&~\cite{RyschkaMarek1981}\\
 "                    &48.78(7)       & 0.1(7)       &VC&HQB+DD&~\cite{Yei1998}\\
 "                    &48.78(7)       & 0.(7)         &-&-&~Recommended\\
5$^2D_{5/2}$ &-21.2(2)        & 0.0(10)     &AB&LIF&~\cite{FredrikssonSvanberg1980}\\
"                      &-22.1(5)       &0.               &VC&ORFDR&~\cite{LamHapper1980}\\ 
"                      &-21.24(5)     &0.2(5)         &VC&HQB+DD&~\cite{Yei1998} \\
"                      &-21.24(5)     &0.2(5)         &-&-&~Recommended\\
5$^2F_{5/2}$ &<|0.7|            &0.               &VC&ORFDR&~\cite{SvanbergHapper1973}\\
5$^2F_{7/2}$ &<|1.0|            &                  &"&"&~"\\
6$^2P_{1/2}$ &291.90(12)    &-                &VC&ORFDR&~\cite{Abele1975a} \\
"                     &291.3(7)        &-                 &AB&ORFDR&~\cite{Coc1987} \\
"                     &291.885(80)   & -               &AB&LIF&~\cite{RafacTanner1997}\\
"                     &291.922(20)   & -               &VC&SAS+FCS&~\cite{Udem1999}\\
"                     &291.918(8)     & -               &VC&SAS&~\cite{DasNatarajan2006b}\\
"                     &291.9135(15)  & -               &VC&SAS&~\cite{DasNatarajan2006}\\
"                     &291.9309(12)  & -               &AB&LIF +FC     &~\cite{Gerginov2006}\\
"                     &291.929(1)      & -               &VC&OS      &~\cite{TruongLuiten2015}\\
"                     &291.9263(25)    &-                &-&-&~w.a.(w.e.e.)\\
6$^2P_{3/2}$&50.15(8)            &-1.35(80)       &AB&HOPF&~\cite{ThibaultHuber1981_2}  \\
"                     &50.275(3)         &-0.53(2)        &AB&LIF&~\cite{TannerWieman1988}  \\
"                     &50.28827(23)   &-0.4934(17)  &AB&LIF&~\cite{GerginovTanner2003} \\
"                     &50.28163(86)   &-0.5266(57)  &VC &CCS& ~\cite{DasNatarajan2005} \\
"                     &50.2878(11)     &-0.496(6)      &-&-&~w.a.(w.e.e.)\\
6$^2D_{3/2}$&16.30(15)         &0.                 &VC &TPSDS&~\cite{TaiGupta1975} \\
"                     &16.17(17)         &0.11(127)     &VC &TPSDS&~\cite{OhtsukaSuzuki2005} \\
"                     &16.34(3)           & -0.1(2)         &VC &OODR&~\cite{Kortyna2006} \\
"                     &16.3331(80)     & -0.36(1)       &VC &OODR&~\cite{ChengHsu2017} \\
"                     &16.338(3)         & -0.136(24)   &VC &TPSDS&~\cite{ChenCheng2018} \\
"                     &16.338(3)         &-0.136(24)    &-&-&~See text. Recommended\\
6$^2D_{5/2}$ &-4.69(4)            & 0.18(73)       &MOT&TPSDS&~\cite{GeorgiadesKimble1994} \\
"                     &-4.56(9)             &-0.35(183)    &VC &TPSDS&~\cite{OhtsukaSuzuki2005} \\
"                     &-4.66(4)             & 0.9(8)          &VC&OODR&~\cite{Kortyna2006} \\
"                     &-4.59(6)             & -0.78(66)     &VC&OODR& ~\cite{WangYang2020} \\
"                     &-4.629(14)         & -0.10(15)     &VC&TPSDS&~\cite{HerdWilliams2021} \\
"                     &-4.629(11)         &-0.10(15)       &-&-&  ~Recommended\\
7$^2S_{1/2}$ &545.90(9)          & -                  &AB &LIF&~\cite{GilbertWieman1983}\\
"                     &545.818(16)     & -                  &VC&OODR&~\cite{YangWang2016,RenWang2016}\\
"                     &545.90(32)       & -                  &MOT    &TPSDS&~\cite{TianZhang2019}\\
"                     &545.87(1)         & -                  &VC    &OODR+EIT&~\cite{HeZhao2020}\\
"                     &545.856(14)       &-                  &-&-&~w.a.(w.e.e.)\\
7$^2P_{1/2}$ &94.35(4)           &-                  &VC     &ORFDR& ~\cite{FeiertagzuPutlitz1972}\\
"                     &94.2(5)             &-                  &VC     &SAS& ~\cite{GerhardtTImmermann1978}\\
"                     &94.40(5)           &-                  &VC     &SAS& ~\cite{WilliamsHawkins2018}\\
"                     &94.37(3)           &-                  &-&-&~w.a.\\
7$^2P_{3/2}$ &16.605(6)         &-0.15(3)      & VC &VR& From~\cite{Arimondo1977} \\
"                     &16.6(3)             &0.                &VC &HQB&~\cite{DeechSeries1977} \\
"                     &16.605(6)         &-0.19(5)       &VC     &SAS&~\cite{WilliamsHawkins2018}\\
"                     &16.605(4)        &-0.16(3)        &-&-&~w.a.\\
7$^2D_{3/2}$ &7.36(3)             & -0.1(2)        &AB&OODR&~\cite{Kortyna2008} \\
"                     &7.386(15)          &-0.18(16)     &VC&OODR+FC&~\cite{StalnakerTanner2010}\\ 
"                     &7.36(7)              & -0.88(87)    &VC&TPSDS&~\cite{LeeChui2011} \\
"                     &7.38(1)              & -0.18(10)    &VC&TPSDS&~\cite{KiranKumar2013}\\
"                     &7.38(19)            &-0.15(21)     &VC&OODR+FC&~\cite{JinJia2013} \\
"                     &7.39(6)              & -0.19(18)    &VC &TPSDS&~\cite{WangJia2021} \\
"                      &7.380(8)           &-0.17(7)        &-&-&~w.a.\\
7$^2D_{5/2}$&-1.56(9)              &0.                 &VC &ELC&~\cite{Auzinsh2007} \\
"                     & -1.717(15)          &-0.18(52)      &VC&OODR+FC&~\cite{StalnakerTanner2010}\\
"                     &-1.81(5)               & 1.01(106)    &VC &TPSDS&~\cite{LeeChui2011} \\
"                     &-1.70(3)               & -0.77(58)    &VC &OODR+EIT&~\cite{WangWang2020} \\
"                     &-1.79(5)               & 1.05(29)      &VC.  &TPSDS&~\cite{WangJia2021}\\
"                     &-1.717(15)           &-0.18(52)      &-&-&~See text. Recommended.\\
8$^2S_{1/2}$ &225.(15)              &-                   &VC  &LOF&~\cite{CampaniPolacco1978} \\
"                     &219.3(2)               &-                   &TD  &TPSDS&~\cite{HerrmannWeis1985} \\
 "                    &219.(1)                 &-                   &MOT&OODR&~\cite{FortTinoInguscio1995} \\
 "                    &220.(1)                 &-                   &VC   &TPSDS&~\cite{FortInguscioSasso1995} \\
 "                    &205.(15)               &-                   &VC   &FC+QBS&~\cite{BelliniHaensch1997} \\
"                     &219.12(1)             &-                   &VC    &TPSDS&~\cite{HagelBiraben1999} \\  
"                     &219.125(4)           &-                   &VC    &TPSDS+FC&~\cite{FendelHaensch2007} \\              
"                     &219.14(11)           &-                   &VC    &OODR+FC&~\cite{StalnakerTanner2010} \\ 
"                     &219.124(7)           &-                   &VC    &SAS&~\cite{WuCheng2013} \\ 
"                     &219.08(12)           &-                   &VC&OODR+EIT&~\cite{WangHe2013,WangWang2014Cs} \\ 
"                     &219.137(11)         &-                    &VC&OODR+FC&~\cite{JinJia2013} \\
"                     &219.125(3)           &-                    &-&-&~w.a.\\
8$^2P_{1/2}$&42.97(10)              &-                &VC     &ORFDR+MFD& ~\cite{TaiHapper1973}\\
"                    & 42.9(3)                 &-                 &VC     &SAS& ~\cite{CataliottiInguscio1996}\\
"                    &42.96(9)                &-                 &-&-&~w.a. \\  
8$^2P_{3/2}$&7.55(5)                  &0.63(35)     &VC  &ORFDR&~\cite{BarbeyGeneux1962} \\
"                    &7.626(5)                  &-0.049(42)     &VC &ORFDR&~\cite{BuckavonOppen1963} \\
"                    &7.58(1)                  &-0.14(5)     &VC &ORFDR&~\cite{FaistKoide1964} \\
                     &7.644(25)              &0.               &VC &ORFDR&~\cite{Abele1975b} \\
"                    &7.42(6)                  &0.14(29)     &VC &HQB+DD&~\cite{Bayram2014} \\
"                    &7.58(3)                  &-0.14(5)      &-&-&~See text. w.a.(w.e.e.);   Recommended.\\  
8$^2D_{3/2}$&3.94(8)                 &0.               &VC &MLC+HQB& From~\cite{Arimondo1977} \\
"                    &3.92(7)                  &0.                &VC&HQB&~\cite{DeechSeries1977} \\
"                    &3.92(10)                &0.                &VC&MFD&~\cite{WijngaardenSagle1991b} \\
"                    &3.95(1)                  &0.                &VC&HQB&~\cite{SagleWijngaarden1991} \\ 
"                    &3.95(1)                  &0.                &-&-&~Recommended\\  
8$^2D_{5/2}$&-0.85(20)               &0.               &VC&ORFDR+MLC&From
~\cite{Arimondo1977}\\
9$^2S_{1/2}$&110.1(5)                 &-                 &VC  &ORFDR&~\cite{Farley1977}\\ 
"                    &109.93(9)               &-                 &VC &OODR+FC&~\cite{StalnakerTanner2010} \\ 
"                    &109.7(3)                 &-                 & VC&TPSDS&~\cite{KiranKumar2012}\\
"                    &110.150(13)           &-                 &VC&OODR+FC&~\cite{JinJia2013} \\
"                    &109.999(3)             &-                 &VC&TPSDS+FC &~\cite{MorgenwegEikema2014} \\
"                    &110.999(3)             &-                 &-&-&~ Recommended \\  
9$^2P_{1/2}$ &23.19(15)              &-                &VC &ORFDR& ~\cite{TsekerisFarleyGupta1975} \\
9$^2P_{3/2}$ &4.123(3)                &-0.051(25)   &VC&MLC&  ~\cite{RydbergSvanberg1972}\\ 
9$^2D_{3/2}$ &2.35(4)                 &0.              &VC&MLC+HQB&From~\cite{Arimondo1977} \\
"                      &2.32(4)                &0.             &VC &HQB&~\cite{DeechSeries1977} \\
"                      &2.38(1)                &0.              &VC&HQB& ~\cite{SagleWijngaarden1991} \\  
"                      &2.375(10)                 &0.                &-&-&~w.a.\\  
9$^2D_{5/2}$&-0.43(4)            & 0.             &VC        &ELC&\cite{Auzinsh2007} \\
10$^2S_{1/2}$& 63.2(3)          &-                &VC&ORFDR&~\cite{TsekerisSvanberg1974}\\ 
10$^2P_{1/2}$& 13.9(2)          &-                &VC&ORFDR&~\cite{Farley1977} \\
10$^2P_{3/2}$& 2.481(9)        &-0.025(8)   &VC&MLC& ~\cite{RydbergSvanberg1972} \\   
10$^2D_{3/2}$&1.52(3)           &0.               &VC  &MLC& ~\cite{SvanbergTsekeris1975}\\
"                      &1.51(2)           &0.               &VC  &HQB& ~\cite{DeechSeries1977}\\
"                      & 1.54(2)          &0.               &VC     &HQB& ~\cite{SagleWijngaarden1991} \\
"                      &1.503(91)       &0.               &VC   &TPSDS&~\cite{OttoHuettner2002} \\
"                      &1.524(13)           &0.                 &-&-&~w.a.\\  
10$^2D_{5/2}$&-0.34(3)         &0.                &VC&ELC&\cite{Auzinsh2007} \\
11$^2S_{1/2}$&39.4(2)           &-                  &VC   &ORFDR&~\cite{TsekerisSvanberg1974}\\
"                       &39.4(17)         &-                  &VC  &TPSDS&~\cite{OttoHuettner2002} \\
"                       &38.81(23)       &-                  & VC&OODR+EIT&\cite{HeWhang2012} \\ 
"                       &39.15(15)       &-                  &-&-&~w.a. (w.e.e.)\\  
11$^2P_{3/2}$&1.600(15)        &0.                &VC &ORFDR&~\cite{BelinSvanberg1974}\\                    
11$^2D_{3/2}$&1.055(15)        &0.                &VC &MLC&~\cite{SvanbergBelin1974} \\
"                      &1.05(4)            &0.                 &       VC &HQB& ~\cite{DeechSeries1977} \\
"                      &1.11(11)          &0.                 &      VC  &TPSDS&~\cite{OttoHuettner2002} \\
"                      &1.0530(69)      &0.                 &      VC  &TPSDS+FC&~\cite{QuirkElliott2022} \\
"                      &1.0530(69)          &0.                 &-&-&~Recommended
\\
11$^2D_{5/2}$&-0.24(6)           &0.                 &VC&ORFDR&~\cite{SvanbergBelin1974}  \\ 
"                      &-0.21(6)           &0.                 &VC  &TPSDS+FC&~\cite{QuirkElliott2022} \\
"                      &-0.225(41)            &0.                &-&-&~w.a.\\
12$^2S_{1/2}$&26.31(10)         &-                  &VC&ORFDR&~\cite{TsekerisGupta1975} \\   
"                      &26.4(16)           &-                   &VC &TPSDS&~\cite{OttoHuettner2002} \\
"                      &26.318(15)       &-                   &VC &TPSDS+FC&~\cite{QuirkElliott2022} \\ 
"                      &26.318(15)         &-                    &-&-&~Recommemded\\
12$^2P_{3/2}$&1.10(3)             &0.                  &VC &ORFDR& ~\cite{BelinSvanberg1976b}\\
12$^2D_{3/2}$&0.758(12)         &0.                  &VC &MLC& ~\cite{SvanbergBelin1974}  \\
"                      &0.75(2)              &0.                  &              VC &HQB& \cite{DeechSeries1977}  \\
"                      &0.758(12)           &0.                   &-&-&~Recommended\\
12$^2D_{5/2}$&-0.19(5)            &0.         &VC &ORFDR& ~\cite{SvanbergBelin1974}  \\
13$^2S_{1/2}$&18.4(1)              &-          &VC  &ORFDR&~\cite{Farley1977} \\ 
"                      &18.6(18)            &-           &VC &TPSDS&~\cite{OttoHuettner2002} \\
"                      &18.431(10)        &-           &VC &TPSDS+FC&~\cite{QuirkElliott2022} \\
"                      &18.431(10)        &-.          &-&-&~Recommended\\  
13$^2P_{3/2}$&0.77(5)              &0.          &VC &ORFDR& ~\cite{BelinSvanberg1976b}\\
13$^2D_{3/2}$&0.556(8)            &0.          &VC &MLC& ~\cite{SvanbergBelin1974}  \\
"                      &0.55(4)               &0.         &VC &HQB& \cite{DeechSeries1977} \\
"                      &0.556(8)             &0.         &-&-&~Recommemded\\  
13$^2D_{5/2}$&-0.14(4)             &0.          &VC &ORFDR& ~\cite{SvanbergBelin1974} \\
14$^2S_{1/2}$&13.4(1)               &-            &VC&ORFDR&~\cite{Farley1977}\\
"                      &13.9(15)             &-             &VC   &TPSDS&~\cite{OttoHuettner2002} \\ 
"                      &13.4(1)               &-             &-&-&~Recommended\\ 
14$^2D_{3/2}$&0.425(7)            &0.            &      VC  &MLC& ~\cite{BelinSvanberg1976b} \\  
"                      &0.40(5)               &0.            &       VC  &HQB& ~\cite{DeechSeries1977} \\ 
"                     &0.425(7)              &0.            &-&-&~Recommended\\ 
15$^2S_{1/2}$&10.1(1)                &-           &           "VC  &ORFDR&~\cite{Farley1977} \\ 
15$^2D_{3/2}$&0.325(8)             &0.          &                  VC  &MLC& ~\cite{BelinSvanberg1976b} \\ 
"                      &0.31(2)               &0.         &VC  &HQB& ~\cite{NakayamaSeries1981} \\ 
"                      &0.325(8)             &0.          &-&-&~Recommended\\ 
16$^2S_{1/2}$&7.73(5)               &-           &VC &ORFDR&~\cite{Farley1977} \\
16$^2D_{3/2}$&0.255(12)          &0.         &VC &MLC&~\cite{BelinSvanberg1976b} \\ 
 "                      &0.24(2)              &0.         &VC &HQB&~\cite{NakayamaSeries1981} \\
 "                      &0.255(12)          &0.                   &-&-&Recommended\\   
 17$^2S_{1/2}$&6.06(10)        &-           &VC &ORFDR&~\cite{Farley1977} \\ 
 17$^2D_{3/2}$&0.190(12)      &0.         &VC &MLC&~\cite{BelinSvanberg1976b}  \\ 
 18$^2D_{3/2}$&0.160(10)      &0.         &"&"&" \\ 
 23$^2S_{1/2}$ & 2.3(2)          &-           &AB   &OODR+MWS &~\cite{RaimondHaroche1981}, also in \cite{GoyHaroche1982} \\ 
 23$^2P_{1/2}$ & 0.56(5)        &-            &    "   &"       &~" \\ 
 25$^2S_{1/2}$ & 1.5(2)          &-            & "      &"        &~" \\ 
 25$^2P_{1/2}$ & 0.40(5)        &-            & "      &"        &~" \\ 
 26$^2S_{1/2}$ & 1.4(2)          &-            & "      &"        &~" \\ 
 26$^2P_{1/2}$ &0.31(5)         &-             &  "     &"       &~" \\ 
 28$^2S_{1/2}$ &1.2(2)           &-             &   "    &"       &"\\ 
 28$^2P_{1/2}$ &0.28(5)         &-             &"       &"       &~"\\
 45$^2P_{3/2}$ &0.0103(27)   &0.            &MOT &MWS&\cite{Sassmannshausen2013} \\ 
 49$^2S_{1/2}$ &0.147(4)       &-              & "       &"       &~" \\
 59$^2P_{3/2}$ &0.0047(10)   &0.            &   "     &"      &~"\\
 67$^2P_{3/2}$ &0.0030(17)   &"              &   "     &"&~"\\ 
 68$^2S_{1/2}$ &0.0520(13)   &-              &   "     &"&~"\\
 72$^2P_{3/2}$ &0.0021(36)   &0.             &   "   &"&~"\\ 
81$^2S_{1/2}$ &0.0318(19)    &-               &   "   &"&~"\\ 
90$^2S_{1/2}$ &0.0227(28     &-                &   "   &"&~"\\ 
66$^2D_{3/2}$ &0.0026(5)     &0.               &"     &"&~" \\ 
66$^2D_{5/2}$ &0.00010(45) &0.               &"    &"   &~"\\
\end{longtable*}

\subsection{Francium}
In the 1980-87 years, the Orsay group at the ISOLDE atomic beam facility in CERN  measured several hyperfine constants of different Fr isotopes reported in~\cite{Liberman1980,Coc1985,Coc1987,Duong1987}. Francium spectroscopy has reached a higher precision level  with the preparation of a cold atom MOT in 1996-97 by~\cite{SimsarianVoytas1996,Lu1997}. Two years later, the important information on the nuclear structure associated with the $^2P_{1/2}$ hyperfine structure pointed out by ~\cite{Grossman1999} has triggered a new interest leading to several more recent experimental investigations.\\
\indent Table~\ref{Table:Fr} reports  all results for the Fr nuclear ground state configuration, avoiding duplicates when the same value was published more than once. The Table does not include the francium data published by~\cite{Voss2013}, where systematic errors are unaccounted, since updates and corrections of these measurements were presented by~\cite{Voss2015}; see Table I of that paper. Table ~\ref{Table:Fr} evidences that a very large set of isotopes was investigated, all of them targeted at the nuclear structure exploration.  The agreement between different hyperfine values is not exceptional,  in several cases the differences being greater than the error bars. 
Because for most isotopes the explored energy levels are limited in number, usually only the lower levels for each $L$ series, global analyses are not efficient. A partial spectroscopic information is associated with the $S$ series in the 210 and 212 isotopes composed by three explored states,  with both $n=8$ and $n=9$ hyperfine values missed out in the second isotope. The inconsistency in~\cite{GomezSafronova2008} between the number in Table 1 and that reported in the text and the abstract, is resolved by using the number reported in the text and the abstract as checked against the original data by one of us (LAO). 
\begin{longtable*}{cddccl}
\caption{Measured $A$ and $B$ values for Fr isotopes. We use g next to the isotope to indicate the ground state of the nucleus. \label{Table:Fr}}\\
\hline
State& A (MHz) & B(MHz)&Sample&Technique  &Ref.   \\
\hline
\endfirsthead
\multicolumn{6}{c}%
{\tablename\ \thetable\ -- \textit{Continued from previous page}} \\
\hline
State& A (MHz)& B(MHz)&Sample&Technique  &Ref. \\
\hline
\endhead
\hline \multicolumn{6}{r}{\textit{Continued on next page}} \\
\endfoot
\hline
\endlastfoot
&&&{\bf $^{202g}$Fr}&\\ 
7$^2S_{1/2}$ &12800.(50)      &-               &AB&LIF&~\cite{Flanagan2013}, also in \cite{Lynch2014}\\
\hline
&&&{\bf $^{203g}$Fr}&\\ 
7$^2S_{1/2}$ &8180.(30)         &-                   &AB&RIS&~\cite{Lynch2014}\\
"                      &8187.(2)          &-                  &AB&LIF&~\cite{Wilkins2017}\\
"                     &8187.(2)    &-                &-&-&~w.a.\\
8$^2P_{3/2}$ &29.5(2)             &-39.1(20)     &"&"&~\cite{Wilkins2017}\\
\hline
&&&{\bf $^{204g}$Fr}&\\
7$^2S_{1/2}$ &12990.(30)      &-                   &AB&RIS&~\cite{Lynch2014}\\
"                     &13146.7(7)      &-                   &AB&LIF&~\cite{Voss2015} \\
"                     &13146.6(36)    &-                   &-&-&~w.a. (w.e.e)\\
7$^2P_{3/2}$ &141.7(3)          &-37.1(19)       &"&"&~\cite{Voss2015}\\

\hline
&&&{\bf $^{205g}$Fr}&\\
7$^2S_{1/2}$ &8355.0(11)      &-                    &AB&LIF&~\cite{Voss2013,Voss2015}\\
"                     &8400.(30)        &-                    &AB&RIS&~\cite{Lynch2014}\\
"                     &8355.0(11)      &-                     &-&-&~Recommended\\
7$^2P_{3/2}$ &89.7(4)            &-81.0(48)       &AB&LIF&~\cite{Voss2015}\\
\hline
&&&{\bf $^{206g}$Fr}&\\
7$^2S_{1/2}$ &13052.2(18)    &-                     &"&"&~\cite{Voss2015} \\
 "                    &13057.8(10)     & -                   &AB&RIS&~\cite{Lynch2016}\\
 "                     &13056.5(24)    &-                    &-&-&~w.a.(w.e.e.)\\
7$^2P_{1/2}$ &1716.90(16)     &-                    &MOT&FML&~\cite{ZhangSprouse2015}\\
7$^2P_{3/2}$ &139.1(8)           &-66.8(50)     &AB&LIF&~\cite{Voss2015}\\
8$^2P_{3/2}$ &47.5(10)           &-29.8(10)      &AB&LIF&~\cite{Lynch2016}\\
\hline
&&&{\bf $^{207g}$Fr}&\\
7$^2S_{1/2}$ &8484.(1)         &-                 &AB&HOPF&~\cite{Coc1985}\\
"                     &8480.(30)       &-                 &AB&LIF&~\cite{Lynch2014}\\
"                     &8482.(2)         &-                 &AB&LIF&~\cite{Wilkins2017}\\
"                     &8483.6(9)    &-                &-&-&~w.a.\\
7$^2P_{1/2}$ &1111.81(11)    &-                 &MOT&FML&~\cite{ZhangSprouse2015}\\
7$^2P_{3/2}$ &90.7(6)           &-42.(13)      &AB&HOPF&~\cite{Coc1985}\\
8$^2P_{3/2}$ &30.4(2)           &-20.0(16)     &AB&LIF&~\cite{Wilkins2017}\\
\hline
&&&{\bf $^{208g}$Fr}&\\ 
7$^2S_{1/2}$ &6639.7(70)         &-              &AB&HOPF&~\cite{Liberman1980}\\
"                     &6650.7(8)           &-              &AB&HOPF&~\cite{Coc1985}\\
"                     &6653.7(4)           &-              &AB&LIF&~\cite{Voss2015}\\
"                     &6653.1(10)    &-                &-&-&~w.a.(w.e.e.)\\
7$^2P_{1/2}$ &874.8(3)             &-               &MOT&FML&~\cite{Grossman1999}\\
7$^2P_{3/2}$ &72.8(5)               &8.9(75)     &AB&HOPF&~\cite{Liberman1980}\\
"                     &72.4(5)                &1.(10)        &AB&HOPF&~\cite{Coc1985}\\
"                     &71.9(2)               &13.6(29)   &AB&LIF&~\cite{Voss2015}\\
"                     &72.1(2)    &6.9(59)                &-&-&~w.a.\\
\hline
&&&{\bf $^{209g}$Fr}&\\
7$^2S_{1/2}$&8590.5(105)   &-               &AB&HOPF&~\cite{Liberman1980}\\ 
"                &8606.7(9)          &-               &AB&HOPF&~\cite{Coc1985}\\
"                     &8606.6(9)    &-                &-&-&~w.a.\\
7$^2P_{1/2}$&1127.9(2)       &-               &MOT&FML&~\cite{Grossman1999}\\
"                    &1127.67(11)       &-               &MOT&FML&~\cite{ZhangSprouse2015}\\
"                    &1127.72(10)    &-                &-&-&~w.a.\\
7$^2P_{3/2}$&93.1(6)           &-61.0(58)  &AB&HOPF&~\cite{Liberman1980}\\
"                &93.3(5)              &-62.(5)       &AB&HOPF&~\cite{Coc1985}\\
"                     &93.2(4))    &-62.(4)               &-&-&~w.a.\\
7$^2D_{5/2}$&-21.(1)            &-81.(22)  &MOT&LIF&~\cite{Agustsson2017}\\
\hline
&&&{\bf $^{210g}$Fr}&\\
7$^2S_{1/2}$&7182.4(81)   &-                 &AB&HOPF&~\cite{Liberman1980}\\
"                &7195.1(4)         &-                   &AB&HOPF&~\cite{Coc1985}\\
"                     &7195.1(6)    &-                &-&-&~w.a.\\
7$^2P_{1/2}$ &945.6(58)    &-                  &AB&HOPF&~\cite{Coc1987}\\
"                &946.3(2)           &-                   &MOT&FML&~\cite{Grossman1999}\\
"                     &946.3(2)    &-                &-&-&~w.a.\\
7$^2P_{3/2}$&77.9(2)         &47.6(22)       &AB&HOPF&~\cite{Liberman1980}\\
"                    &78.0(2)            &51.(4)           &AB&HOPF&~\cite{Coc1985}\\
"                    &77.95(14)    &48.4(19)                &-&-&~w.a.\\
7$^2D_{3/2}$&22.3(5)         &0.                 &MOT&OODR&~\cite{Grossman2000}\\
7$^2D_{5/2}$ &-17.8(8)       &64.(17)         &"&"&"\\
8$^2S_{1/2}$&1577.8(11)   &-                   &MOT&TCSDS&\cite{Simsarian1999}\\
9$^2S_{1/2}$&622.25(36)   &-                   &MOT&TCSDS&\cite{GomezSafronova2008}\\
\hline
&&&{\bf $^{211g}$Fr}&\\
7$^2S_{1/2}$&8698.2(105)   &-                 &AB&HOPF&~\cite{Liberman1980}\\ 
"                 &8713.9(8)          &-                  &AB&HOPF&~\cite{Coc1985}\\
"                 &8700.(60)           &-                  &AB&RIS&~\cite{Lynch2014} \\
"                     &8713.8(8)    &-                &-&-&~w.a.\\
7$^2P_{1/2}$ &1142.1(2)      &-                 &MOT&FML&~\cite{Grossman1999}\\
7$^2P_{3/2}$&94.7(2)           &-55.3(34)   &AB&HOPF&~\cite{Liberman1980}\\ 
"                     &94.9(3)               &-51.(7)        &AB&HOPF&~\cite{Coc1985}\\
"                     &94.8(2)    &-54.5(31)               &-&-&~w.a.\\
\hline
&&&$^{212g}$Fr&\\ 
7$^2S_{1/2}$&9051.3(95)     &-                 &AB&HOPF&~\cite{Liberman1980}\\ 
"                 &9064.2(2)           &-                 &AB&HOPF&~\cite{Coc1985}\\
"                 &9064.4(15)         &-                &AB&LIF&~\cite{Duong1987}\\
"                     &9064.2(2)    &-                &-&-&w.a.\\
7$^2P_{1/2}$ &1189.1(46)     &-                &AB&HOPF&~\cite{Coc1987}\\
"                 &1187.1(68)          &-                &AB&LIF&~\cite{Duong1987}\\
"                 &1192.0(2)            &-                &MOT&FML&~\cite{Grossman1999}\\
"                     &1192.0(2)    &-                &-&-&~w.a.\\
7$^2P_{3/2}$&99.1(9)            &-35.3(155)  &AB&HOPF&~\cite{Liberman1980}\\ 
"                &97.2(1)                &-26.(2)        &AB&LIF&~\cite{Coc1985}\\
"                &97.2(1)                &-26.0(2)      &AB&LIF&~\cite{Duong1987}\\
"                     &97.21(10)    &-26.0(2)                &-&-&~w.a.(w.e.e.); w.a.\\
8$^2P_{1/2}$&373.0(1)          &-                &AB&LIF&~\cite{Duong1987}\\
8$^2P_{3/2}$&32.8(1)            &-7.7(9)      &AB&"&"\\
8$^2D_{3/2}$ &13.0(6)           &0.             &AB&LIF&~\cite{Arnold1990}\\
8$^2D_{5/2}$&-7.2(6)             &0.             &AB&"&"\\
9$^2D_{3/2}$ &7.1(7)             &0.             &AB&"&"\\
9$^2D_{5/2}$ &-3.6(4)            &0.             &AB&"&"\\
10$^2S_{1/2}$ &401.(5)         &-              &AB&"&"\\
11$^2S_{1/2}$ &225.(3)         &-              &AB&"&"\\
\hline
&&&$^{213g}$Fr&\\ 
7$^2S_{1/2}$&8744.9(105)     &-                 &AB&HOPF&~\cite{Liberman1980}\\ 
"                &8759.9(6)             &-                  &AB&HOPF&~\cite{Coc1985}\\
"                &8757.4(19)           &-                  &AB&LIF&~\cite{Duong1987}\\
"                     &8759.6(6)    &-                &-&-&~w.a.\\
7$^2P_{1/2}$&1150.5(75)       &-                  &AB&HOPF&~\cite{Coc1987}\\
"                &1147.89(11)          &-                  &MOT&FML&~\cite{ZhangSprouse2015}\\
"                     &1147.89(11)    &-                &-&-&~Recommended\\
7$^2P_{3/2}$&94.5(16)           &-20.7(170)   &AB&HOPF&~\cite{Liberman1980}\\ 
"                 &95.3(3)                &-36.(5)          &AB&HOPF&~\cite{Coc1985}\\
"                 &95.3(3)                &-36.0(5)        &AB&LIF&~\cite{Duong1987}\\
"                     &95.3(2)    &-36.0(5)              &-&-&~w.a.\\
8$^2P_{3/2}$&31.6(1)            &-7.0(25)         &AB&LIF&~\cite{Duong1987}\\
\hline
&&&$^{214g}$Fr&\\
7$^2S_{1/2}$&2370.(150)        &-                        &AB&RIS&~\cite{Faroqsmith2016,FaroqSmith2016_Err}\\  
\hline
&&&$^{219g}$Fr&\\ 
7$^2S_{1/2}$ &6820.(30)         &-                    &AB&RIS&~\cite{Budincevic2014}\\
"                      &6851.(1)          &-                    &AB&RIS&~\cite{deGroote2015}\\
"                     &6851.(1)           &-                &-&-&Recommended\\
8$^2P_{3/2}$&24.7(5)              &-104.(1)          &"&"&~\cite{deGroote2015}\\
\hline
&&&$^{220g}$Fr&\\ 
7$^2S_{1/2}$&-6549.4(9)       &-                  &AB&HOPF&~\cite{Coc1985}\\ 
"                    &-6549.2(12)      &-                  &AB&LIF&~\cite{Duong1987}\\
"                    &-6500.(40)         &-                  &AB&RIS&~\cite{Lynch2014} \\
"                     &-6549.3(7)    &-                &-&-&~w.a.\\
7$^2P_{3/2}$&-73.2(5)           &126.8(5)      &AB&HOPF&~\cite{Coc1985}\\
"                 &-68.5(62)            &123.(9)        &AB&HOPF&~\cite{Coc1987}\\
"                     &-73.2(5)    &125.9(44)                &-&-&~w.a.\\
8$^2P_{3/2}$&-23.3(1)           &41.4(14)      &AB&LIF&~\cite{Duong1987}\\
\hline
&&&$^{221g}$Fr&\\  
7$^2S_{1/2}$&6204.6(8)        &-                 &AB&HOPF&~\cite{Coc1985}\\
"                    &6100.(200)      &-                 &AB&HOPF&~\cite{AndreevLethokov1986}\\
"                    &6205.6(17)      &-                 &AB&HOPF&~\cite{Coc1987}\\
"                    &6209.9(10)      &-                &AB&LIF&~\cite{Duong1987}\\
"                    &6200.(30)         &-                  &AB&RIS&~\cite{Lynch2014} \\
"                    &6209.(1)           &-               &AB&RIS&~\cite{deGroote2015}\\
"                     &6207.2(11)    &-                &-&-&~w.a.(w.e.e.)\\
7$^2P_{1/2}$&808.(12)          &-               &AB&HOPF&~\cite{Coc1987}\\
"                    &811.0(13)          &-                &MOT&LIF&~\cite{Lu1997}\\
"                 &810.3(18)          &-                &MOT& FML&~\cite{ZhangSprouse2015}\\
"                     &810.7(10)    &-                &-&-&~w.a.\\
7$^2P_{3/2}$&65.5(6)          &-264.(3)   &AB&HOPF&~\cite{Coc1985}\\
"                &65.4(29)            &-259.(16)    &AB&HOPF&~\cite{Coc1987}\\
"                &66.5(9)              &-260.(48)   &MOT&LIF&~\cite{Lu1997}\\
"                     &65.8(49)       &-264.(3)                &-&-&~w.a.\\
8$^2P_{3/2}$&22.4(1)          &-85.7(8)    &AB&LIF&~\cite{Duong1987}\\
"                    &22.3(5)          &-87.(2)      &AB&RIS&~\cite{deGroote2015}\\
"                    &22.40(10)      &-86.9(3)                &-&-&~w.a.; w.a.(w.e.e.)\\
\hline
&&&$^{222g}$Fr&\\
7$^2S_{1/2}$&3070.(3)        &-              &AB&HOPF&~\cite{Coc1985}\\
7$^2P_{3/2}$&33.(1)             &133.(9)     &"  &"                 &"\\
\hline
&&&$^{223g}$Fr&\\
7$^2S_{1/2}$&7654.(2)        &-               &"&"&"\\
7$^2P_{3/2}$&83.3(9)          &308.(3)      &"   &"                 &"\\
\hline
&&&$^{224g}$Fr&\\ 
7$^2S_{1/2}$&3876.(1)         &-                &"&"&"\\
7$^2P_{3/2}$&42.1(7)           &136.(1)      &"   &"                  &"\\
\hline
&&&$^{225g}$Fr&\\
7$^2S_{1/2}$ &6980.(1)         &-               &"    &"                  &"\\
"                     &6980.1(75)         &-                &AB&HOPF&~\cite{Coc1987}\\
"                     &6980.(1)    &-                &-&-&Recommended\\
7$^2P_{3/2}$&77.1(5)           &347.(2)       &AB&HOPF&~\cite{Coc1985}\\
"                 &77.2(30)            &346.(13)     &AB&HOPF&~\cite{Coc1987}\\
"                     &77.1(5)    &347.(2)                &-&-&~w.a.\\
\hline
&&&$^{226g}$Fr&\\
7$^2S_{1/2}$&699.4              &-                 &AB&HOPF&~\cite{Coc1985}\\
                      &698.1071(20) &-                 &"   &"                 &~\cite{Duong1986}\\
                      "                     &698.107(2)    &-                &-&-&~w.a.\\
7$^2P_{3/2}$&7.(1)                   &-356.(4)  &"    &"                 &~\cite{Coc1985}\\
\hline
&&&$^{227g}$Fr&\\
7$^2S_{1/2}$&29458.(4)            &-                &"    &"                  &"\\
7$^2P_{3/2}$&316.(2)             &0.    &"&"&"\\
\hline
&&&$^{228g}$Fr&\\
7$^2S_{1/2}$&-3731.(4)          &-                &"      &"                  &"\\
7$^2P_{3/2}$&-41.(2)              &627.(12)     &"      &"                  &"\\
\hline
&&&$^{229g}$Fr&\\
7$^2S_{1/2}$&30080.(110)     &-                &AB&RIS&~\cite{Budincevic2014}\\
\hline
&&&$^{231g}$Fr&\\
7$^2S_{1/2}$&30770.(130)      &-              &"&"&"\\
\end{longtable*}

\section{Data analysis}
\subsection{Quantum number scaling law}
\label{Scalinglaws}
\begin{figure}[!!t]
\centering
\includegraphics[width=9 cm]{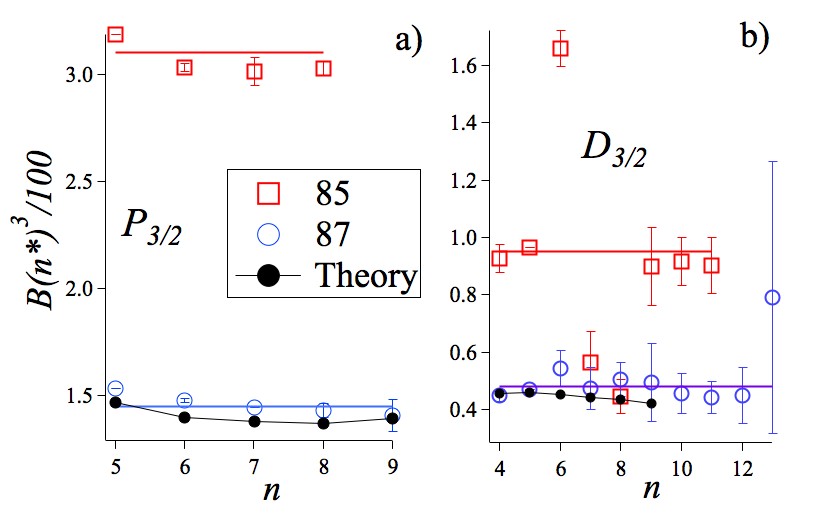}
\caption{$B(n^*)^3$ scaling, with $B$ in MHz, for Rb isotopes versus $n$, with open blue circles for  the $^{87}$Rb data, and open red squares for the $^{85}$Rb data, with error bars for the experiments. (a) $^2P_{3/2}$ data; (b) $^2D_{3/2}$ data. The  continuous horizontal lines  represent fits based on the  $(n^*)^{-3}$ scaling law. For the $^{85}$Rb  $^2D_{3/2}$ data,  the fit does not include the $n=6, 7$ states. The black dots joined by a line represent the theoretical predictions by~\cite{SafronovaSafronova2011} for $^{87}$Rb. }
\label{ScalingBPDstates}
\end{figure}

\begin{figure}[!!t]
\centering
\includegraphics[width=7 cm]{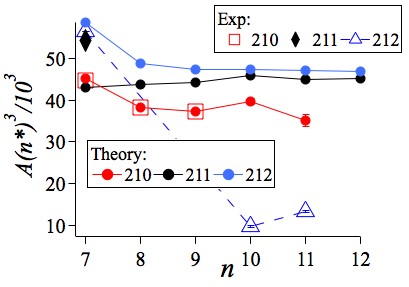}
\caption{$A(n^*)^3$ scaling, with $A$ in MHz, vs. $n$ for the 210, 211 and 212 Fr isotopes.  Experimental data: red open squares for $^{210}$Fr, black diamond for the $^{211}$Fr single value, and blue open triangles for $^{212}$Fr, joined by a dashed line. Theoretical predictions with colored dots joined by a continuous line for $^{210}$Fr by~\cite{SahooSakemi2015}, for $^{211}$Fr  by~\cite{GrunefeldGinges2019}, and for $^{212}$Fr by~\cite{LouTang2019}. Error bars for experiment and theory are too small to be  visible.  The effective quantum numbers are derived from  the quantum defects of~\cite{Simsarian1999}. The $^{211}$Fr theoretical data follow closely a horizontal line of the $1/(n^*)^{3}$ scaling law.   }
\label{ScalingAFr}
\end{figure}

The present level of high precision for the theoretical computations leads to agreement with selected experimental results up to 0.1 percent, also for high quantum numbers. Even at that precision level,  semi-empirical laws, such as the scaling ones, remain useful for verifying or predicting data, or confirming the presence of perturbations for specific atomic states.  \\ 
 \begin{figure}
\centering
\includegraphics[width= 1.0\columnwidth]{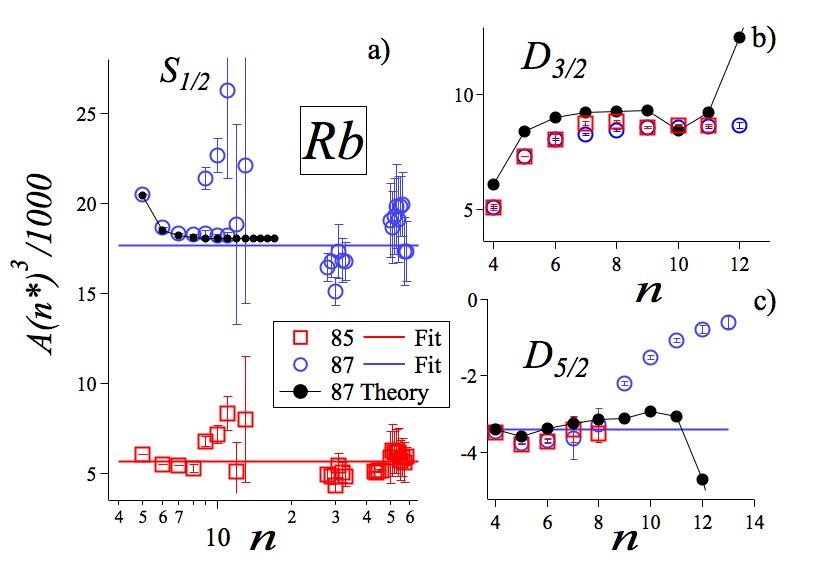}
\includegraphics[width= 1.0\columnwidth]{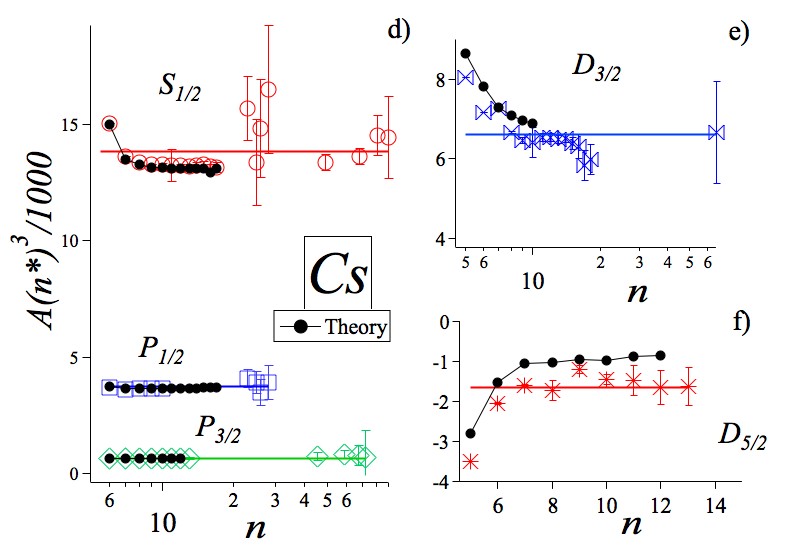}
\caption{ $A(n^*)^3$ scaling test, with $A$ in MHz,  vs. $n$ number. Experimental result  with their error bars are shown in colors, while black dots represent the theoretical predictions. Experimental and theoretical data:  (a) data for  Rb $^2S_{1/2}$ states, and  (b) and (c) for $^2D_{3/2,5/2}$ ones, with red open squares for $^{85}$Rb and blue circles for $^{87}$Rb; (d) data for Cs $^2S_{1/2}$ and $^2P_{1/2,3/2}$,  (e) and (f) for Cs $^2D_{3/2}$ and $^2D_{5/2}$, respectively. Panel (f) plot does not include the $n=66$ value of Table~\ref{Table:Cs} owing to its very large error bar. $^{85}$Rb  $^2$D states data are scaled to the $^{87}$Rb ones by assuming the validity of the isotope $g_I$ scaling of Eq.~\eqref{Ascaling}.    Note  the logarithmic horizontal scale in (a), (d) and (e).  The continuous horizontal lines represent fits based on the $1/(n^*)^3$  scaling. Theoretical predictions are: for $^{87}$Rb $^2S_{1/2}$ by~\cite{GrunefeldGinges2019} and $^2D_{3/2,5/2}$  by \cite{SafronovaSafronova2011}; for Cs  $^2S_{1/2}$ and $^2P_{1/2}$ states by~\cite{GrunefeldGinges2019}, $^2P_{3/2}$ and $^2D_{3/2,5/2}$   by~\cite{TangShi2019}; for the Cs $^2D_{5/2}$ states, the data by ~\cite{Auzinsh2007} appear superimposed. }
\label{AScalingValid}
\end{figure}
\indent We have tested the scaling laws of Eq.~\eqref{Ascaling} for $A$ and similar one of $B$, for the overall $S$, $P$ and $D$ states of potassium, rubidium, cesium and francium, using the quantum defect parameters from ~\cite{LorenzenNiemax1983,LiGallagher2003, LorenzenNiemax1984,Simsarian1999,PeperDeiglmayr2019}.  As examples, we report in Fig.~\ref{AScalingValid} the $A$ dipole constant results for both Rb isotopes, in (a) for $^2S$ states, and in (b), (c) for the $^2D$ ones.   Panel (d) of that figure reports the data for the $^2S$ and $^2P$ states of Cs, and for the $^2D$ ones.  The $^2S$  $n=(12-13)$ $^{87}$Rb data by~\cite{StoicheffWeinberger1979} with large error bars are not plotted. Note that the $A(n^*)^3$ values are plotted vs. $n$ enhancing the deviations from the scaling law. The validity of the scaling law is tested by the horizontal lines derived from data fits. For the $^2D$  states in (b) and (c), the $^{85}$Rb data have been scaled to the $^{87}$Rb ones by supposing the validity of the $g_J$ scaling of Eq.~\eqref{gIscaling}, leading to a precise superposition of the two isotope values. Similar results are obtained for all the $^{39}K$ states and the $^2S$ states of $^{41}K$.  \\
\indent For the  $^2S$ states, the $^{87}$Rb theoretical results by~\cite{GrunefeldGinges2019} reproduce very closely the experimental $A$ values for all quantum numbers, as shown by the black dots in the (a) panel of the figure. The $A$ scaling law was tested theoretically for the $^2S$, $^2P$ and $^2D$ Rb states in the log/log plot of~\cite{SafronovaSafronova2011}. In panels (b) and (c) of Fig.~\ref{AScalingValid}, the black dots depict those theoretical predictions for the $^2D_{3/2}$ states.   For the values up to $n\approx 9$, the differences between theoretical and experimental results are small. For higher $n$ values, the differences are significant, because of the limitation in the computer codes  at that time. A good agreement with experimental data exists for the ~\cite{GrunefeldGinges2019} predictions of the Cs $^2S_{1/2}$ and $^2P_{1/2}$ states, as shown in (d).  For Cs, there are additional theoretical results for the $^2P_{3/2}$ states by~\cite{TangShi2019}, for $^2D_{3/2}$  by~\cite{Auzinsh2007}, and for $^2D_{5/2}$ by~\cite{TangShi2019}. For the $^2D$  states, the  data by~\cite{Auzinsh2007} cannot be distinguished on the figure scale. In addition, they cover a short range of $n$ values.  \\
\indent In most cases, the $A$ scaling law is verified in both experimental and theoretical data, and its validity is used to assign the sign of the $A$ values reported in the Tables of the previous Section.  For the Cs $^2P_{1/2,3/2}$ and $^2D_{3/2,5/2}$ states in a large range of $n$ values, the horizontal fits  are  good. The scaling validity applies to the high-$n$ states, as for the $(n=40,90)$ $^2S_{1/2}$ and $^2P_{3/2}$ data of~\cite{Sassmannshausen2013} as seen in panel (d). For the $^{87}$Rb and Cs $^2S$ states, the scaling  does not apply precisely to low-$n$ values because of additional contributions to $A$ in  Eq.~\eqref{Aconstant}.  In contrast, the $^{85}$Rb values satisfy the $1/(n^*)^3$ scaling. The low-$n$ difference between the two isotopes produces the deviation from the $g_J$  scaling,  linked to the hyperfine anomalies discussed in the following subsection. For the low-$n$ $^2D$ states, large deviations from the scaling lead to lower $A$ values in Rb  and higher $A$ ones in Cs for both experimental and theoretical data. These deviations are equivalent for the two Rb  isotopes. They originate from the pair-correlation and core-polarization, as explained in~\cite{Auzinsh2007,TangShi2019}. \\  
\indent We have verified the validity of the $1/(n^*)^{3}$  scaling law  also for  the $B$ constants.  Fig.~\ref{ScalingBPDstates} reports the $B(n^*)^{3}$ constants of both  Rb isotopes for the $^2P_{3/2}$ and $^2D_{3/2}$ states. The continuous horizontal  lines indicate that the scaling law is valid  for the  $^{87}$Rb $B$ constants of both states. For the $^{85}$Rb isotope, deviations appear  in Fig~\ref{ScalingBPDstates}(b) for the $^2D_{3/2}$ states at intermediate $n$ values. The 
$B$ theoretical predictions by~\cite{SafronovaSafronova2011} denoted by black dots indicate a good agreement between the theory and experiments. The $B$ scaling applies also to the Cs $^2P_{3/2}$ states. For the Rb $^2D_{5/2}$ and Cs $^2D$ states, a definitive conclusion cannot be reached owing to the limited number of data. For the $^2P_{3/2}$ states, the $^{85}$Rb $B$ data are precisely scaled to the $^{87}$Rb ones by assuming the validity of the quadrupole moment dependence of Eq.~\eqref{Qscaling}, with the $Q$ values of~\cite{Raghavan1989}. This $Q$ proportionality applies also to the $^2D_{3/2}$ data, where the $n^*$  scaling is valid.\\
\indent Since the data for each Fr isotope are very limited in number, the application of the quantum number scaling law is not  very efficient. Neverthless, its validity test is useful to verify the present level of knowledge. The law is also useful for the experimental determination of the hyperfine constants in states not yet explored. While the francium effective quantum numbers may be derived  from the quantum defects of~\cite{Arnold1990,Simsarian1999,HuangSun2010}, we rely on those of~\cite{Simsarian1999} based on the full spectrum of the  francium absorption lines compiled by~\cite{Sansonetti2007}. The francium Table evidences that hyperfine sequences are available only for the $^2S$ states of the 210 and 212 isotopes, each sequence limited to three entries. Fig.~\ref{ScalingAFr} presents  those experimental results in a $A(n^*)^3$ plot. The dipole constant for the 211 isotope 7$^2S$ state  is also plotted. The dots joined by continuous lines represent the theoretical data for $^{210}$Fr by~\cite{SahooSakemi2015}, for $^{211}$Fr by~\cite{GrunefeldGinges2019}, and for $^{212}$Fr by~\cite{LouTang2019}. On the figure scale, equivalent results are obtained using the quantum defect numbers of all the above references.  For the experimental and theoretical data on the $^{210,212}$Fr isotopes, the deviations from the scaling law for the lowest $n=7$ number are similar to those presented for the Rb and Cs low-$n$ $^2S$ and $^2D$ states in Fig.~\ref{AScalingValid}. Such deviation from the scaling law does not appear in the $^{211}$Fr theoretical data by~\cite{GrunefeldGinges2019}, while their prediction for the $^2S$ Cs and Rb states  match very closely the experimental data as shown in Fig.~\ref{AScalingValid}.  An acquisition of more experimental data is  required in order to progress with this exploration. The program highlighted by~\cite{GrunefeldGinges2019} and by~\cite{RobertsGingers2020} focuses on the importance of having more hyperfine splitting information for higher excited levels to better understand not only the Fr atom, but the properties of the different nuclear isotopes.     
\subsection{Anomalies}
\label{Anomalies}
\begin{figure}
\centering
\includegraphics[width=8 cm]{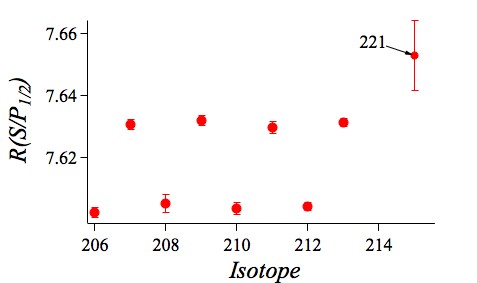}
\caption{$R(S/P_{1/2})$ ratio for the $7^2S_{1/2}$ and $7^2P_{1/2}$ states of francium isotopes vs the isotope number. Mean values and error bars are derived from the francium Table~\ref{Table:Fr}. Notice the staggered dependence  of $R(S/P_{1/2})$ on the even/odd isotope number. The $^{221}$Fr isotope value with a large error bar agrees with that dependence.}
\label{Ratio_Fr}
\end{figure}

The hyperfine anomalies,  first introduced by \cite{BohrWeisskopf1950}, are defined as the $A$ deviations from Eq.~\eqref{gIscaling} produced by the finite size of the nucleus.  As a measure of the finite structure influence on the dipole constants of isotopes 1 and 2, following \cite{Persson2013} the expression for the $^1\Delta^2$ hyperfine anomaly is:
\begin{equation}
^1\Delta^2= \frac{A^1}{A^2}\frac{g_I^2}{g_I^1}-1,
\label{anomaly}
\end{equation}
where $(A^i,g^i_{I})$ are the hyperfine magnetic dipole constant and the nuclear gyromagnetic ratio, respectively,  of the $i=(1,2)$ isotopes.   For a point-like nucleus, the hyperfine anomaly is null.                                         
The article of ~\cite{Persson2013} represents the most recent review of the atomic anomalies. Those of the $6^2S_{1/2}$ Fr states are  examined in~\cite{ZhangSprouse2015}. Recent theoretical studies of Fr anomalies can be found  in~\cite{Konovalova2018,Konovalova2020,RobertsGingers2020}.\\
\indent Anomalies can be derived from measured $A$ constants  and accurate values of the nuclear $g$-ratio. This is the case for light alkali atoms with their precise values in the gas phase determined  in the sixties and seventies, as discussed in~\cite{Arimondo1977}. Anomalies for those atoms are reported in Table~\ref{AnomaliesLight}, derived from the weighted mean values and variances of the dipole constants reported in the Tables of the previous Section. Only anomalies with a value significantly different from zero are presented in Table~\ref{AnomaliesLight}. For the $5^2P_{1/2}$ state of the Rb isotopes, where a very large discrepancy between the measured values was noted in subsection~\ref{measuredconstants}, the  value is missed because none reasonable anomaly is associated to the different set of data.    The Rb values of Table~\ref{AnomaliesLight} are in good agreement with  those derived by~\cite{PerezGalvanSprouse2007,PerezGalvanOrozco2008,WangWang2014Rb} for $^2S$ and $^2D$ states. For the Rydberg $S$ states, enormous anomalies are obtained, $\approx -50(5)$ percent, a quite surprising result, because the interaction of a Rydberg electron with the nucleus should be comparable to that of low  orbitals. The fairly constant anomaly for all the $^2S$ states of the Rb isotopes was pointed out by \cite{PerezGalvanSprouse2007}. For Rb $^2P$ states the situation is not well defined, with the the $5^2P_{1/2}$ value reflecting the discording results associated to this state, and  for the $^2P_{3/2}$ states the quadrupole interaction playing an important role.  The constant value of the $4^2P_{1/2}$ anomaly applies also to the K isotopes.\\
\begin{table} 
\caption{Light alkali hyperfine anomalies $^1\Delta^2$ with states listed in order of increasing $L$, then of increasing $n$ and finally of increasing $J$ }
\label{AnomaliesLight}
\begin{tabular}{ccccc}
\hline
\hline
 Element 
&Isotope 1& Isotope 2& State& $^1\Delta^2(\%)$\\
\hline
 Li  &  6&7   &2$S_{1/2}$ &0.0068067(8)\\ 
     &     &     &2$P_{1/2}$ &-0.1734(2) \\
     &     &     &2$P_{3/2}$ &-0.155(8)    \\
 K  &39  &40 &4$S_{1/2}$ &0.467(2)\\
     &     &      &4$P_{1/2}$ & 3.90(13)\\ 
K  & 39&41 &4$S_{1/2}$ &-0.22937(13)\\
     &      &     &4$P_{1/2}$ & 3.9(3)\\ 
Rb & 85 & 87&5$S_{1/2}$ &0.35141(2)\\
      &     &     &6$S_{1/2}$ &0.361(19)\\
      &     &     &7$S_{1/2}$ &0.342(3)\\
      &     &     &5$P_{1/2}$ &0.55(8)\\
      &     &     &5$P_{3/2}$ &0.168(5)\\
      &     &     &6$P_{1/2}$ &0.31(7)\\
      &     &     &6$P_{3/2}$ &0.46(5)\\
      &     &     &4$D_{5/2}$ &0.60(15)\\
      &     &     &5$D_{3/2}$ &0.279(6)\\
      &     &     &5$D_{5/2}$ &0.44(5)\\
 \hline
\end{tabular}
\end{table}
\indent The francium case is different, because the list of isotope data is quite long and therefore, interesting information about the nuclear structures could be derived from the anomaly determinations. However, for francium an  important element is missing in Eq.~\eqref{anomaly} because a direct measurement of the nuclear gyromagnetic ratio is available only for $^{211}$Fr in~\cite{Stone2005}. For the remaining isotopes, the $g_I$ ratios are derived by assuming a zero anomaly: see~\cite{Raghavan1989}. In the recent theoretical studies of Fr anomalies by~\cite{Konovalova2018,RobertsGingers2020,Konovalova2020}, the information on the nuclear structure  is replaced by derivations of the radial nuclear structure and of the nuclear radius. In order to obtain  the anomalies without relying on such theoretical analyses, ~\cite{Grossman1999},  following \cite{Persson1998}, concentrated their attention on the $7^2S_{1/2}$ and $7^2P_{1/2}$ states.
The $^2P_{1/2}$ electron probes the nucleus with a more uniform radial dependence of the interaction than does the $^2S_{1/2}$ electron. They introduced the following $R(S/P)$ ratio of their hyperfine constants:
\begin{equation}
R(S/P)=\frac{A(S_{1/2})}{A(P_{1/2})}
\end{equation} 
as a probe of the nuclear magnetization distribution.  Since both states are spin-1/2, these ratios are independent of quadrupole effects that complicate the extraction of the nuclear structure information. For several francium isotopes, the ratio is  presented in Fig~\ref{Ratio_Fr}. Notice the staggered isotope behaviour of $R(S/P)$ vs. the even/odd  isotope number.  Such behavior evidences the different radial distributions of the nuclear magnetization for the even/odd number of neutrons. The ratios of the hyperfine constants of the $7^2S_{1/2}$ and $7^2P_{3/2}$ states, as well as those of the $7^2P_{1/2}$ and $7^2P_{3/2}$  ones, do not exhibit such a clear staggered dependence as the $R(S/P)$ ratio plotted in the figure. We have tested the presence of a $R(S/P)$  staggered dependence also for the Rb isotopes using the 85 and 87 data in Table~\ref{Table:85Rb},\ref{Table:87Rb}, and the 82 isotope ones from~\cite{ZhaoVieira1999}.  This limited data set appears to confirm  the $R(S/P)$ staggered behavior.\\
\indent     The ratio of the hyperfine constants of the $^2P_{1/2}$  and $^2P_{3/2}$  states has been proposed in~\cite{Konovalova2020} as an additional test of the nuclear structure, even if the nuclear quadrupole coupling is important for the latter state. This ratio calculated from the francium Table~\ref{Table:Fr} data does not exhibit a clear dependence on the isotope number. \\

\section{Conclusions}
The previously published experimental values for the stable isotopes of the light alkali atoms and for all the nuclear-ground-configuration francium isotopes have been compiled. The Tables report the most accurate data obtained before 1977 and all those published  after that time. For each measured hyperfine constant we present a recommended value. A critical examination of the most interesting cases, or of the most discording ones,  is presented. For the  discording cases we have calculated a weighted enhanced error
following the procedure of the Particle Data Group in \cite{PDG}.   For those cases a comparison of our data analysis to the cluster maximum likelihood estimator introduced by \cite{Rukhin2009,Rukhin2019}, is presented in the supplementary information (SI) of this review~\cite{SI_Hyperfine}.  We encourage future reviewers of larges sets of data to look into these more recent methods, that are currently gaining adepts in the community, but are not followed by those in charge of the recent redefinition of the fundamental constants of physics and chemistry \cite{Tiesinga2021}.\\
\indent The experimental interest in measuring  hyperfine constants is renewed in the recent years. Today, the laser sources required to excite energy levels not easily assessed at the optical pumping time are available on the market. Those sources and the associated atomic species can be important for quantum simulation and computational investigations. As in the past, the quest for higher precision spectroscopic measurement could represent an additional step for refining the experimental tools required in those areas.\\
\indent Instead of concentrating the attention on a specific atomic state, the recent global theoretical analyses cover all the atomic states of a single species, possibly of all the isotopes. For this global approach, precise experimental data are required for a large set of states,  stimulating the hyperfine data search in several directions.\\
\indent Spectroscopic investigations of alkali atoms are also associated with the theoretical progress on the electron-nucleus hyperfine coupling.  The hyperfine coupling being an important probe of the nuclear structure, the ~\cite{Grossman1999} francium research has introduced the ratio of the $S$ and $P$ state hyperfine couplings as a new tool for studying nuclear properties. This  has stimulated a large theoretical effort, but it has pointed out the need for more precise experimental data. The application of this ratio to lighter alkalis could be an interesting exploration direction.  \\
\indent  The electron-nucleus interaction contains a term characterized by the nuclear magnetic octupole moment. In recent years three different measures of that moment are reported for $^{133}$Cs  $6^2P_{3/2}$ by ~\cite{GerginovTanner2003},  for $^{87}$Rb $5^2P_{3/2}$  in the experimental data of ~\cite{YeHall1996}   reexamined by ~\cite{Gerginov2009}, and more recently for the $^{133}$Cs $6^2D_{3/2}$ state by ~\cite{ChenCheng2018}. The parity violation search in alkali atoms may also find new life. Up to now, the most accurate data on the atomic parity-non-conserving interaction was derived from the $6^2S_{1/2}\to 7^2S_{1/2}$ transition in cesium by ~\cite{WoodWieman1997}. To obtain
a more accurate value of the nuclear weak charge producing the parity-violating Hamiltonian, it would be desirable to consider other candidates.~\cite{Gwinner2022} with their collaborators are pursuing its measurement in the $7^2S_{1/2}\to 8^2S_{1/2}$ transition in a variety of Fr isotopes, while ~\cite{AokiSakemi2017} have proposed to search for that violation operating on the $^7S_{1/2}\to 6^2D_{3/2}$ electric quadrupole transition in $^{210}$Fr. It is expected that advances in the understanding of hyperfine interactions will continue to illuminate atomic parity nonconversion and vice versa.\\
\indent As a final expectation, in the near future, a few totally new entries will be added to the above Tables, counterbalanced by several refined entries.

\section{Acknowledgments}
The authors are grateful to Marianna Safronova for a guide through the recent progress in the theoretical analyses of the hyperfine structures and a careful reading of the manuscript, and to Roberto Calabrese for hints on the francium spectroscopy.   The authors are grateful and indebted to the referee for ample comments that have made this a much better review and for providing the algorithm of the CMLE approach. Hassan Jawahery, Peter J. Mohr and William D. Phillips have helped elucidate the evaluation of the errors to report recommended values.  The authors acknowledge Wang-Yau Cheng, Pierre Dub\'e,  Johannes  Deiglmayr, Randy J. Knize, Mark Lindsay, Frederic Merkt, Dieter Meschede, Priyanka Rupasinghe and Sun Svanberg for communicating data of various atomic states. The very kind help of the librarians Armelle Michetti and Odile Richaud in Grenoble, and Massimiliano Bertelli in Pisa is also acknowledged.

\bibliography{biblioHyperfineSurvey4}

\end{document}